\documentstyle [10pt,amsfonts] {article}
\input epsf

\newcommand{\beq}{\begin{equation}}
\newcommand{\eeq}{\end{equation}}
\newcommand{\beqs}{\begin{eqnarray}}
\newcommand{\eeqs}{\end{eqnarray}}

\catcode`@=11
\@addtoreset{equation}{section}
\@addtoreset{equation}{subsection}
\def\theequation{\ifnum\value{section}=0 \arabic{equation}\ignorespaces
\else \ifnum\value{section}=-1 A.\arabic{equation}\ignorespaces
\else \ifnum\value{subsection}=0 \thesection.\arabic{equation}\ignorespaces
\else \thesection.\arabic{subsection}.\arabic{equation}\ignorespaces
                           \fi
                      \fi
                 \fi}
\catcode`@=12

\begin{document}

\def\thefootnote{\fnsymbol{footnote}}

\baselineskip 5.0mm

\vspace{4mm}

\begin{center}

{\Large \bf Exact Partition Function for the Potts Model with Next-Nearest
Neighbor Couplings on Strips of the Square Lattice}

\vspace{8mm}

\setcounter{footnote}{0}
Shu-Chiuan Chang$^{(a)}$\footnote{email: shu-chiuan.chang@sunysb.edu} and
\setcounter{footnote}{6}
Robert Shrock$^{(a,b)}$\footnote{(a): permanent address;
email: robert.shrock@sunysb.edu}

\vspace{6mm}

(a) \ C. N. Yang Institute for Theoretical Physics  \\
State University of New York       \\
Stony Brook, N. Y. 11794-3840  \\

(b) \ Physics Department \\
Brookhaven National Laboratory \\
Upton, NY  11973

\vspace{10mm}

{\bf Abstract}
\end{center}

We present exact calculations of partition function $Z$ of the $q$-state Potts
model with next-nearest-neighbor spin-spin couplings, both for the
ferromagnetic and antiferromagnetic case, for arbitrary temperature, on
$n$-vertex strip graphs of width $L_y=2$ of the square lattice with free,
cyclic, and M\"obius longitudinal boundary conditions.  The free energy is
calculated exactly for the infinite-length limit of these strip graphs and the
thermodynamics is discussed.  Considering the full generalization to arbitrary
complex $q$ and temperature, we determine the singular locus ${\cal B}$ in the
corresponding ${\mathbb C}^2$ space, arising as the accumulation set of
partition function zeros as $n \to \infty$.

\vspace{16mm}

\pagestyle{empty}
\newpage

\pagestyle{plain}
\pagenumbering{arabic}
\renewcommand{\thefootnote}{\arabic{footnote}}
\setcounter{footnote}{0}

\section{Introduction}

The $q$-state Potts model has served as a valuable model for the study of phase
transitions and critical phenomena \cite{potts,wurev}.  On a lattice, or, more
generally, on a (connected) graph $G$, at temperature $T$, this model is
defined by the partition function
\beq
Z(G,q,v) = \sum_{ \{ \sigma_n \} } e^{-\beta {\cal H}}
\label{zfun}
\eeq
with the (zero-field) Hamiltonian
\beq
{\cal H} = -J \sum_{\langle i j \rangle} \delta_{\sigma_i \sigma_j}
\label{ham}
\eeq
where $\sigma_i=1,...,q$ are the spin variables on each vertex $i \in G$;
$\beta = (k_BT)^{-1}$; and $\langle i j \rangle$ denotes pairs of adjacent
vertices.  The graph $G=G(V,E)$ is defined by its vertex set $V$ and its edge
(=bond) 
set $E$; we denote the number of vertices of $G$ as $n=n(G)=|V|$ and the
number of edges of $G$ as $e(G)=|E|$.  We use the notation
\beq
K = \beta J \ , \quad a = u^{-1} = e^K \ , \quad v = a-1
\label{kdef}
\eeq
so that the physical ranges are (i) $a \ge 1$, i.e., $v \ge 0$ corresponding to
$\infty \ge T \ge 0$ for the Potts ferromagnet, and (ii) $0 \le a \le 1$,
i.e., $-1 \le v \le 0$, corresponding to $0 \le T \le \infty$ for the Potts
antiferromagnet.  One defines the (reduced) free energy per site $f=-\beta F$,
where $F$ is the actual free energy, via
\beq
f(\{G\},q,v) = \lim_{n \to \infty} \ln [ Z(G,q,v)^{1/n}]  \ .
\label{ef}
\eeq
where we use the symbol $\{G\}$ to denote $\lim_{n \to \infty}G$ for a given
family of graphs.

Let $G^\prime=(V,E^\prime)$ be a spanning subgraph of $G$, i.e. a subgraph
having the same vertex set $V$ and an edge set $E^\prime \subseteq E$. Then
$Z(G,q,v)$ can be written as the sum \cite{birk}-\cite{kf}
\beqs
Z(G,q,v) & = & \sum_{G^\prime \subseteq G} q^{k(G^\prime)}v^{e(G^\prime)}
\label{cluster} \cr\cr\cr
& = & \sum_{r=k(G)}^{n(G)}\sum_{s=0}^{e(G)}z_{rs} q^r v^s
\label{zpol}
\eeqs
where $k(G^\prime)$ denotes the number of connected components of $G^\prime$
and $z_{rs} \ge 0$.  Since we only consider connected graphs $G$, we have
$k(G)=1$. The formula (\ref{cluster}) enables one to generalize $q$
from ${\mathbb Z}_+$ to ${\mathbb R}_+$ (keeping $v$ in its physical range).
This generalization is sometimes denoted the random cluster model \cite{kf};
here we shall use the term ``Potts model'' to include both positive integral
$q$ as in the original formulation in eqs. (\ref{zfun}) and (\ref{ham}), and
the generalization to real (or complex) $q$, via eq. (\ref{cluster}).  The
formula (\ref{cluster}) shows that $Z(G,q,v)$ is a polynomial in $q$ and $v$
(equivalently, $a$) with maximum and minimum degrees indicated in
eq. (\ref{zpol}).  The Potts model partition function on
a graph $G$ is essentially equivalent to the Tutte polynomial
\cite{tutte1}-\cite{tutte5} and Whitney rank polynomial \cite{whit},
\cite{wurev}, \cite{bbook}-\cite{boll} for this graph, as was discussed in
\cite{a} and is briefly noted in the appendix of this paper. 

\vspace{6mm}

In this paper we shall present exact calculations of the Potts model partition
function for strips of the square lattice with next-nearest-neighbor (NNN)
spin-spin interactions.  Specifically, we consider strips with width $L_y=2$
vertices and arbitrarily great length $L_x$ with various boundary conditions.
To avoid increasing the number of parameters, we take the nearest-neighbor and
next-nearest-neighbor coupling strengths to be equal.  This study is a natural
continuation of our previous analogous calculations for strips of the square
and triangular lattices \cite{bcc,a,ta}, and the reader is referred to these
papers for background and further references (see also \cite{1dnnn,ks}).  
We envision
the strip as being formed by starting with a ladder graph, i.e. a $2 \times
L_x$ strip of the square lattice, and then adding an edge joining the lower
left and upper right vertices of each square, and an edge joining the upper
left and lower right vertices of each square (these two added edges do not
intersect each other).  Following our previous papers \cite{w3,k}, we shall
denote this lattice as $sq_d$, where the subscript $d$ refers to the added
diagonal bonds for each square.  The Potts model with NNN spin-spin couplings
on the square lattice and strips thereof can equivalently be considered as the
Potts model with nearest-neighbor couplings on the $sq_d$ lattice with these
diagonal bonds present.  For our strip calculations, we take the longitudinal
(transverse) direction on the strip to be the horizontal, $x$ (vertical, $y$)
direction, respectively.  We use free transverse boundary conditions and
consider free, periodic (= cyclic), and M\"obius longitudinal boundary
conditions.  As we showed before \cite{k}, for a given value of $L_x=m$, the
$L_y=2$ cyclic $sq_d$ strip graph is identical to the corresponding strip with
M\"obius boundary conditions; hence, we shall refer to them both as
$L_m=sq_d((L_y=2)_m,FBC_y,(T)PBC_x)$, where the $L$ here stands for ``ladder''
(with diagonal bonds added). The open strip will be denoted $S_m$.  Following
our labelling conventions in \cite{strip,hs}, $L_x=m+1$ edges for an open strip
$S_m$ and $L_x=m$ edges for the cyclic/M\"obius strip $L_m$.  One has
$n(S_m)=2(m+2)$, $n(L_m)=2m$, $e(S_m)=6+5m$, and $e(L_m)=5m$.  Each vertex on
the cyclic/M\"obius strip $L_m$ has degree (coordination number) $\Delta=5$;
this is also true of the interior vertices on the open strip $S_m$, while the
corner vertices have $\Delta=3$.  Hence, the cyclic/M\"obius strips $L_m$ are
$\Delta$-regular graphs with $\Delta=5$, where a $\Delta$-regular graph is
defined as one in which each vertex has the same degree, $\Delta$.  For the
infinite $sq_d$ lattice, $\Delta=8$.  Note that, regarded as graphs, the
cyclic/M\"obius strips of the $sq_d$ lattice considered here, like the full 2D
$sq_d$ lattice, are nonplanar (except for $L_m$ with $m=1,2$).  In contrast,
the open strips $S_m$ of the $sq_d$ lattice are planar graphs.

One interesting special case is provided by the zero-temperature Potts
antiferromagnet.  In general, for sufficiently large $q$, on a given lattice 
or graph $G$, the 
Potts antiferromagnet exhibits nonzero ground state entropy (without
frustration).  This is equivalent to a ground state degeneracy per site
(vertex), $W > 1$, since $S_0 = k_B \ln W$.  The $T=0$ (i.e., $v=-1$) partition
function of the above-mentioned $q$-state Potts antiferromagnet (PAF) on $G$
satisfies
\beq 
Z(G,q,-1)=P(G,q)
\label{zp}
\eeq
where $P(G,q)$ is the chromatic polynomial (in the variable $q$) 
expressing the number
of ways of coloring the vertices of the graph $G$ with $q$ colors such that no
two adjacent vertices have the same color \cite{birk,bbook,rrev,rtrev}. The
minimum number of colors necessary for this coloring is the chromatic number
of $G$, denoted $\chi(G)$.  We have
\beq
W(\{G\},q)= \lim_{n \to \infty}P(G,q)^{1/n} \ . 
\label{w}
\eeq

There are several motivations for the present study.  Clearly, new exact
calculations of Potts model partition functions are of value in their own
right.  A specific motivation is that this study provides exact results that
reveal the effects of next-nearest-neighbor spin-spin interactions on the
properties of the Potts model.  It is of considerable physical interest what
these effects are, since models with strictly nearest-neighbor interactions are
only an approximation (albeit often a good one) to nature.  For ferromagnetic
spin-spin interactions ($J > 0$), the addition of (ferromagnetic) NNN
interactions clearly enhances the tendency, at a given temperature, toward
ferromagnetic ordering.  For antiferromagnetic (AF) spin-spin
interactions ($J < 0$), the effect of the addition of (antiferromagnetic) NNN 
can be investigated by starting with the simple case of zero temperature. 
In the case $q=2$, i.e., the Ising antiferromagnet, if one considers the square
lattice at $T=0$, there is complete antiferromagnetic long range order.
However, in contrast, on the $sq_d$ lattice, the Ising model is
frustrated.  Closely related to this, on the $sq_d$ lattice, the chromatic
number is $\chi(sq_d)=4$, rather than the value 2 for the square (or any
bipartite) lattice.

For a regular lattice, as one increases the lattice coordination number,
the ground state entropy of the $q$-state Potts antiferromagnet (if nonzero for
the given value of $q$), decreases. This can be understood as a consequence of
the fact that as one increases the lattice coordination number, one is
increasing the constraints on the coloring of a given vertex subject to the
condition that other vertices of the lattice adjacent to this one
(i.e. connected with a bond of the lattice) have different colors.  The
addition of NNN spin-spin couplings to the Hamiltonian for the Potts
antiferromagnet on the square lattice has a similar effect of increasing 
the constraints on the values that any given spin can take on, and
hence decreasing the ground state entropy.  
Our exact calculations for the strips
of the square lattice with NNN couplings give a quantitative
measure of this effect.  In a different but related direction, owing to
the correspondence with the Tutte
polynomial, our calculations yield several quantities of relevance to
mathematical graph theory.

Using the formula (\ref{cluster}) for $Z(G,q,v)$, one can generalize $q$ from
${\mathbb Z}_+$ not just to ${\mathbb R}_+$ but to ${\mathbb C}$ and $a$ from
its physical ferromagnetic and antiferromagnetic ranges $1 \le a \le \infty$
and $0 \le a \le 1$ to $a \in {\mathbb C}$.  A subset of the zeros of $Z$ in
the two-complex dimensional space ${\mathbb C}^2$ defined by the pair of
variables $(q,a)$ can form an accumulation set in the $n \to \infty$ limit,
denoted ${\cal B}$, which is the continuous locus of points where the free
energy is nonanalytic.  This locus is determined as the solution to a certain
$\{G\}$-dependent equation in the two complex variables $q$ and $a$
\cite{bcc,a}.  For a given value of $a$, one can consider this locus in the $q$
plane, and we denote it as ${\cal B}_q(\{G\},a)$.  In the special case $a=0$
(i.e., $v=-1$) where the partition function is equal to the chromatic
polynomial, the zeros in $q$ are the chromatic zeros, and ${\cal
B}_q(\{G\},a=0)$ is their continuous accumulation set in the $n \to \infty$
limit \cite{bds}-\cite{read91}. In a series of papers starting with \cite{w} 
we have given exact
calculations of the chromatic polynomials and nonanalytic loci ${\cal B}_q$ for
various families of graphs (for references on this $a=0$ special case, see
\cite{a,ta,s4,k}).  In particular, in \cite{k} we gave an exact determination
of ${\cal B}_q$ for the ($L_x \to \infty$ limit of the) $L_y=2$ $sq_d$ strip.
A motivation for the present study is that it shows how the locus ${\cal
B}_q$ that we calculated for the zero-temperature Potts antiferromagnet
generalizes to finite temperature, as well as to the case of the Potts
ferromagnet. With the exact Potts partition function for arbitrary temperature,
one can study ${\cal B}_q$ for $a \ne 0$ and, for a given value of $q$, one can
study the continuous accumulation set of the zeros of $Z(G,q,v)$ in the $a$
plane; this will be denoted ${\cal B}_a(\{G\},q)$.  It will often be convenient
to consider the equivalent locus in the $u=1/a$ plane, namely ${\cal
B}_u(\{G\},q)$.  We shall sometimes write ${\cal B}_q(\{G\},a)$ simply as
${\cal B}_q$ when $\{G\}$ and $a$ are clear from the context, and similarly
with ${\cal B}_{a}$ and ${\cal B}_{u}$.  One gains a unified understanding of
the separate loci ${\cal B}_q(\{G\})$ for fixed $a$ and ${\cal B}_a(\{G\})$ for
fixed $q$ by relating these as different slices of the locus ${\cal B}$ in the
${\mathbb C}^2$ space defined by $(q,a)$ as we shall do here.

Following the notation in \cite{w} and our other earlier works on ${\cal
B}_q(\{G\})$ for $a=0$, we denote the maximal region in the complex $q$ plane
to which one can analytically continue the function $W(\{G\},q)$ from physical
values where there is nonzero ground state entropy as $R_1$ .  The maximal
value of $q$ where ${\cal B}_q$ intersects the (positive) real axis was
labelled $q_c(\{G\})$.  Thus, region $R_1$ includes the positive real axis for
$q > q_c(\{G\})$.  Correspondingly, in our works on complex-temperature
properties of spin models, we had labelled the complex-temperature extension
(CTE) of the physical paramagnetic phase as (CTE)PM, which will simply be
denoted PM here, the extension being understood, and similarly with
ferromagnetic (FM) and antiferromagnetic (AFM); other complex-temperature
phases, having no overlap with any physical phase, were denoted $O_j$ (for
``other''), with $j$ indexing the particular phase \cite{chisq}.  Here we shall
continue to use this notation for the respective slices of ${\cal B}$ in the
$q$ and $a$ or $u$ planes.  Another motivation for our present study is that it
yields a deeper insight into the singular locus ${\cal B}_q$ for the
zero-temperature Potts antiferromagnet that we found for the $L_x \to \infty$
limit of the $L_y=2$ $sq_d$ strip \cite{k} by showing how this locus changes as
one increases the temperature from zero to finite values.

We note some values of chromatic numbers for the strip graphs considered here:
\beq
\chi(S_m)=4
\label{chism}
\eeq
\beq
\chi(L_m) = \cases{ 4 & for even $m \ge 2$ \cr 
                    5 & for odd  $m \ge 5$ \cr }
\label{chi}
\eeq 
Two degenerate cases are as follows: for $L_x=m=2$, the cyclic strip
reduces to the complete graph\footnote{\footnotesize{The complete graph $K_p$
is the graph with $p$ vertices each of which is adjacent to all of the other
vertices.}} $K_4$, while for $m=3$ it reduces to $K_6$, with $\chi(K_p)=p$.

We record some special values of $Z(G,q,v)$ below.  First, 
\beq
Z(G,q=0,v)=0  \ . 
\label{zq0}
\eeq
This implies that $Z(G,q,v)$ has an overall factor of $q$ and, in general
(and for all the graphs considered here), this is the only overall factor that
it has.  We also have
\beq
Z(G,q=1,v)=\sum_{G^\prime \subseteq G} v^{e(G^\prime)} = a^{e(G)} \ .
\label{zq1}
\eeq
For temperature $T=\infty$, i.e., $v=0$, 
\beq
Z(G,q,v=0)=q^{n(G)} \ .
\label{za1}
\eeq
\beq
Z(G,q,v=-1)=P(G,q)=\biggl [ \prod_{s=0}^{\chi(G)-1}(q-s) \biggr ] U(G,q)
\label{pchi}
\eeq
where $U(G,q)$ is a polynomial in $q$ of degree $n(G)-\chi(G)$. Hence, 
\beq
Z(G,q,v=-1)=P(G,q)=0 \quad {\rm for} \quad G=S_m, \ L_m \quad 
{\rm and} \quad q=1,2,3 \ . 
\label{pq123}
\eeq
The result (\ref{pq123}) implies 
that for $q=2$ and 3, the partition functions 
$Z(G,q=2,v)$ and $Z(G,q=3,v)$ each contain at least one power of the factor 
$(v+1)=a$; for $q=1$, one already knows the form of $Z(G,q=1,v)$ from 
(\ref{zq1}). For the graphs $L_m$ with odd $m$, $\chi(L_m)=5$, so that 
$Z(G,4,v)$ contains at least one power of $(v+1)$ as a factor.

Another basic property, evident from eq. (\ref{cluster}), is that (i) the zeros
of $Z(G,q,v)$ in $q$ for real $v$ and hence also the continuous accumulation
set ${\cal B}_q$ are invariant under the complex conjugation $q \to q^*$; (ii)
the zeros of $Z(G,q,v)$ in $v$ or equivalently $a$ for real $q$ and hence also
the continuous accumulation set ${\cal B}_a$ are invariant under the complex
conjugation $a \to a^*$.

Just as the importance of noncommutative limits was shown in (eq. (1.9) of)
\cite{w} on chromatic polynomials, so also one encounters an analogous
noncommutativity here for the general partition function (\ref{cluster}) of
the Potts model for nonintegral $q$: at certain special points $q_s$
(typically $q_s=0,1...,\chi(G)$) one has
\beq
\lim_{n \to \infty} \lim_{q
\to q_s} Z(G,q,v)^{1/n} \ne \lim_{q \to q_s} \lim_{n \to \infty} Z(G,q,v)^{1/n}
\ .
\label{fnoncomm}
\eeq
Because of
this noncommutativity, the formal definition (\ref{ef}) is, in general,
insufficient to define the free energy $f$ at these special points $q_s$; it is
necessary to specify the order of the limits that one uses in eq.
(\ref{fnoncomm}).  We denote the two
definitions using different orders of limits as $f_{nq}$ and $f_{qn}$:
\beq
f_{nq}(\{G\},q,v) = \lim_{n \to \infty} \lim_{q \to q_s} n^{-1} \ln Z(G,q,v)
\label{fnq}
\eeq
\beq
f_{qn}(\{G\},q,v) = \lim_{q \to q_s} \lim_{n \to \infty} n^{-1} \ln Z(G,q,v) \
.
\label{fqn}
\eeq
In Ref. \cite{w} and our subsequent works on chromatic polynomials and the
above-mentioned zero-temperature antiferromagnetic limit, it was convenient to
use the ordering $W(\{G\},q_s) = \lim_{q \to q_s} \lim_{n \to \infty}
P(G,q)^{1/n}$ since this avoids certain discontinuities in $W$ that would be
present with the opposite order of limits.  In the present work on the full
temperature-dependent Potts model partition function, we shall
consider both orders of limits and comment on the differences where
appropriate.  Of course in discussions of the usual $q$-state Potts model (with
positive integer $q$), one automatically uses the definition in eq.
(\ref{zfun}) with (\ref{ham}) and no issue of orders of limits arises, as it
does in the Potts model with real $q$.
As a consequence of the noncommutativity (\ref{fnoncomm}), it follows that for
the special set of points $q=q_s$ one must distinguish between (i) $({\cal
B}_a(\{G\},q_s))_{nq}$, the continuous accumulation set of the zeros of
$Z(G,q,v)$ obtained by first setting $q=q_s$ and then taking $n \to \infty$,
and (ii) $({\cal B}_a(\{G\},q_s))_{qn}$, the continuous accumulation set of the
zeros of $Z(G,q,v)$ obtained by first taking $n \to \infty$, and then taking $q
\to q_s$.  For these special points,
\beq
({\cal B}_a(\{G\},q_s))_{nq} \ne ({\cal B}_a(\{G\},q_s))_{qn} \ .
\label{bnoncomm}
\eeq
 From eq. (\ref{zq0}), it follows that for any $G$,
\beq
\exp(f_{nq})=0 \quad {\rm for} \quad q=0
\label{fnqq0}
\eeq
and thus
\beq
({\cal B}_a)_{nq} = \emptyset \quad {\rm for} \quad q=0 \ .
\label{bnq0}
\eeq
However, for many families of graphs, including the circuit graph $C_n$,
and cyclic and M\"obius strips of the square or triangular lattice, if we take
$n \to \infty$ first and then $q \to 0$, we find that $({\cal B}_u)_{qn}$ is
nontrivial.  Similarly, from (\ref{zq1}) we have, for any $G$,
\beq
({\cal B}_a)_{nq} = \emptyset \quad {\rm for} \quad q=1
\label{bnq1}
\eeq
since all of the zeros of $Z$ occur at the single discrete point $a=0$ (and in
the case of a graph $G$ with no edges, $Z=1$ with no zeros).  However, as the
simple case of the circuit graph shows \cite{a}, $({\cal B}_u)_{qn}$ is, in
general, nontrivial.

As derived in \cite{a}, a general form for the Potts model partition function
for the strip graphs considered here, or more generally, for recursively
defined families of graphs comprised of $m$ repeated subunits (e.g. the columns
of squares of height $L_y$ vertices that are repeated $L_x$ times to form an
$L_x \times L_y$ strip of a regular lattice with some specified boundary
conditions), is
\beq
Z(G,q,v) = \sum_{j=1}^{N_\lambda} c_{G,j}
(\lambda_{G,j}(q,v))^m
\label{zgsum}
\eeq
where $N_\lambda$ depends on $G$.  This is a generalization of the result of
\cite{bkw} for the chromatic polynomial $P(G,q)$ to the full Potts model
partition function or equivalently, Tutte polynomial. 

The Potts ferromagnet has a zero-temperature phase transition in the $L_x \to
\infty$ limit of the strip graphs considered here, and this has the consequence
that for cyclic and M\"obius boundary conditions, ${\cal B}$ passes through the
$T=0$ point $u=0$.  It follows that ${\cal B}$ is noncompact in the $a$ plane.
Hence, it is usually more convenient to study the slice of ${\cal B}$ in the
$u=1/a$ plane rather than the $a$ plane.  Since $a \to
\infty$ as $T \to 0$ and $Z$ diverges like $a^{e(G)}$ in this limit, we shall
use the reduced partition function $Z_r$ defined by
\beq
Z_r(G,q,v)=a^{-e(G)}Z(G,q,v)=u^{e(G)}Z(G,q,v)
\label{zr}
\eeq
which has the finite limit $Z_r \to 1$ as $T \to 0$.  For a general strip
graph $(G_s)_m$ of type $G_s$ and length $L_x=m$, we can write
\beqs
Z_r((G_s)_m,q,v) & = & u^{e((G_s)_m)}\sum_{j=1}^{N_\lambda} c_{G_s,j}
(\lambda_{G_s,j})^m \equiv \sum_{j=1}^{N_\lambda} c_{G_s,j}
(\lambda_{G_s,j,u})^m
\label{zu}
\eeqs
with
\beq
\lambda_{G_s,j,u}=u^{e((G_s)_m)/m}\lambda_{G_s,j} \ .
\label{lamu}
\eeq
For the $L_y=2$ strips of the $sq_d$ lattice of interest here, the prefactor 
on the right-hand side of (\ref{lamu}) is $u^5$. 

\section{Case of Free Longitudinal Boundary Conditions}

In this section we present the Potts model partition function
$Z(S_m,q,v)$ for the $L_y=2$ strip $S_m$ with arbitrary
length $L_x=m+1$ (i.e., containing $m+1$ squares) and free transverse and
longitudinal boundary conditions.
One convenient way to express the results is in terms of a generating
function:
\beq
\Gamma(S,q,v,z) = \sum_{m=0}^\infty Z(S_m,q,v)z^m \ .
\label{gammazfbc}
\eeq
We have calculated this generating function using the deletion-contraction
theorem for the corresponding Tutte polynomial $T(S_m,x,y)$ and then
expressing the result in terms of the variables $q$ and $v$.  We find
\beq
\Gamma(S,q,v,z) = \frac{ {\cal N}(S,q,v,z)}{{\cal D}(S,q,v,z)}
\label{gammazcalc}
\eeq
where
\beq
{\cal N}(S,q,v,z)=A_{S,0}+A_{S,1}z
\label{numgamma}
\eeq
with
\beq
A_{S,0}=q(16v^3+15qv^2+15v^4+4v^3q+6vq^2+6v^5+v^6+q^3)
\label{as0}
\eeq
\beq
A_{S,1}=-qv^2(v+q)(v+1)(4qv+4v^3q+4v^3+2v^4+2q^2+4vq^2+v^2q^2+11qv^2)
\label{as1}
\eeq
and
\beqs
{\cal D}(S,q,v,z) & = &  1-(12v^2+5qv+q^2+5v^4+10v^3+v^5)z \cr\cr
 & + & v^2(v+1)(2v^4+4v^3+4qv^3+v^2q^2+11qv^2+4vq+4vq^2+2q^2)z^2 \ .
\label{dengamma}
\eeqs
(The generating function for the Tutte polynomial 
$T(S_m,x,y)$ is given in the appendix.) Writing 
\beq
{\cal D}(S,q,v,z) = \prod_{j=1}^2 (1-\lambda_{S,j}z)
\label{ds}
\eeq
we have
\beq
\lambda_{S,(1,2)} = \frac{1}{2}\biggl [ T_{S12} \pm \sqrt{R_{S12}} \ \biggr ]
\label{lams}
\eeq
where
\beq
T_{S12}=12v^2+5qv+q^2+5v^4+10v^3+v^5
\label{t12}
\eeq
and
\beqs
R_{S12} & = & 45v^8+116v^7+196v^6+144v^4+224v^5+104qv^3+40qv^4-4v^3q^2 \cr\cr
& + & 41v^2q^2-6v^6q-10v^5q+q^4-2v^5q^2-10v^4q^2+10vq^3+v^{10}+10v^9
\label{rs12}
\eeqs

In \cite{hs} we presented a formula to obtain the chromatic polynomial for a
recursive family of graphs in the form of powers of $\lambda_j$'s starting
from the
generating function, and the generalization of this to the full Potts model
partition function was given in \cite{a}.  Using this, we have 
\beq
Z(S_m,q,v) = \frac{(A_{S,0}\lambda_{S,1} + A_{S,1})}
{(\lambda_{S,1}-\lambda_{S,2})}\lambda_{S,1}^m +
 \frac{(A_{S,0} \lambda_{S,2} + A_{S,1})}
{(\lambda_{S,2}-\lambda_{S,1})}\lambda_{S,2}^m
\label{pgsumkmax2}
\eeq
(which is symmetric under $\lambda_{S,1} \leftrightarrow \lambda_{S,2}$).
Although both the $\lambda_{S,j}$'s and the coefficient functions involve 
the square root $\sqrt{R_{S12}}$ and are not polynomials in $q$ and $v$,
the theorem on symmetric functions of the roots of an algebraic equation
\cite{pm,uspensky} 
guarantees that $Z(S_m,q,v)$ is a polynomial in $q$ and $v$ (as
it must be by (\ref{cluster}) since the coefficients of the powers of $z$ in
the equation (\ref{dengamma}) defining these $\lambda_{S,j}$'s are
polynomials in
these variables $q$ and $v$).

As will be shown below, in the limit $m \to \infty$ of this strip, the 
singular locus ${\cal B}_u$ consists of arcs that
do not separate the $u$ plane into different regions, so that the PM phase
and its complex-temperature extension occupy all of this plane, except for
these arcs.  For physical temperature and positive integer $q$, the (reduced)
free energy of the Potts model in the limit $n \to \infty$ is given by
\beq
f = \frac{1}{2}\ln \lambda_{S,1} \ .
\label{fstrip}
\eeq
This is analytic for all finite
temperature, for both the ferromagnetic and antiferromagnetic sign of the
spin-spin coupling $J$.  The internal energy and specific heat can be
calculated in a straightforward manner from the free energy
(\ref{fstrip}); since
the resultant expressions are somewhat cumbersome, we do not list them here.
In the $T=0$ Potts antiferromagnet limit $v=-1$,
$\lambda_{S,1}=(q-2)(q-3)$ and $\lambda_{S,2}=0$, so that eq. 
(\ref{gammazcalc})
reduces to the generating function for the chromatic polynomial
for this open strip
\beq
\Gamma(S,q,v=-1;z) = \frac{q(q-1)(q-2)(q-3)}{1-(q-2)(q-3)z} \ . 
\label{gammacp}
\eeq
Equivalently, the chromatic polynomial is
\beq
P(S_m,q) = q(q-1)[(q-2)(q-3)]^{(m+1)} \ .
\label{psq}
\eeq

For the ferromagnetic case with general $q$, in the low-temperature limit
$v \to \infty$,
\beq
\lambda_{S,1} = v^5+5v^4+O(v^3) \, \quad
\lambda_{S,2} = 2v^2 + O(v)  \quad {\rm as} \quad v \to \infty
\label{lamsvinf}
\eeq
so that $|\lambda_{S,1}|$ is never equal to $|\lambda_{S,2}|$ in this
limit, and hence ${\cal B}_u$ does not pass through the origin of the
$u$ plane for the $n \to \infty$ limit of this open $sq_d$ strip:
\beq
u=0 \not\in {\cal B}_u(\{S\}) .
\label{unotinbs}
\eeq
In contrast, as will be shown below, ${\cal B}_u$ does pass through
$u=0$ for this strip with cyclic or M\"obius boundary conditions.

\subsection{${\cal B}_q(\{S\})$ for fixed $a$}

\begin{figure}[hbtp]
\centering
\leavevmode
\epsfxsize=2.5in
\begin{center}
\leavevmode
\epsffile{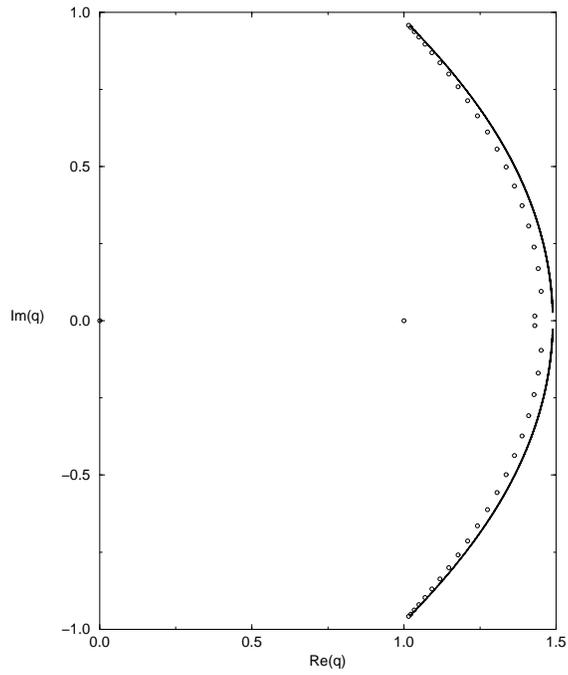}
\end{center}
\caption{\footnotesize{Locus ${\cal B}_q$ for the $n \to \infty$ limit of
the $L_y=2$ $sq_d$ strip $\{S\}$ with free longitudinal boundary 
conditions, for $a=v+1=0.5$. Zeros of $Z(S_m,q,v=-0.5)$ for $m=20$ (i.e.,
$n=44$ vertices, so that $Z$ is a polynomial of degree 44 in $q$) are
shown for comparison.}}
\label{k4xy2a0p5}
\end{figure}

\begin{figure}[hbtp]  
\centering
\leavevmode
\epsfxsize=2.5in
\begin{center}
\leavevmode
\epsffile{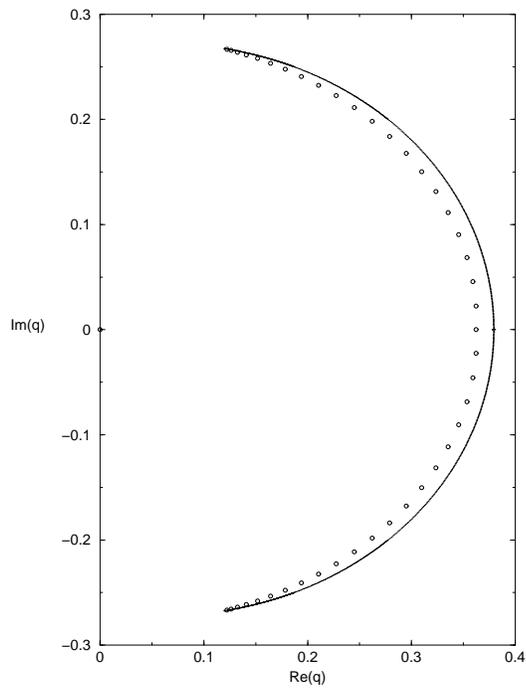}  
\end{center}
\caption{\footnotesize{Locus ${\cal B}_q$: same as Fig. \ref{k4xy2a0p5}
for $a=0.9$ (i.e., $v=-0.1$).}}
\label{k4xy2a0p9}
\end{figure}

\begin{figure}[hbtp]  
\centering
\leavevmode
\epsfxsize=2.5in
\begin{center}
\leavevmode
\epsffile{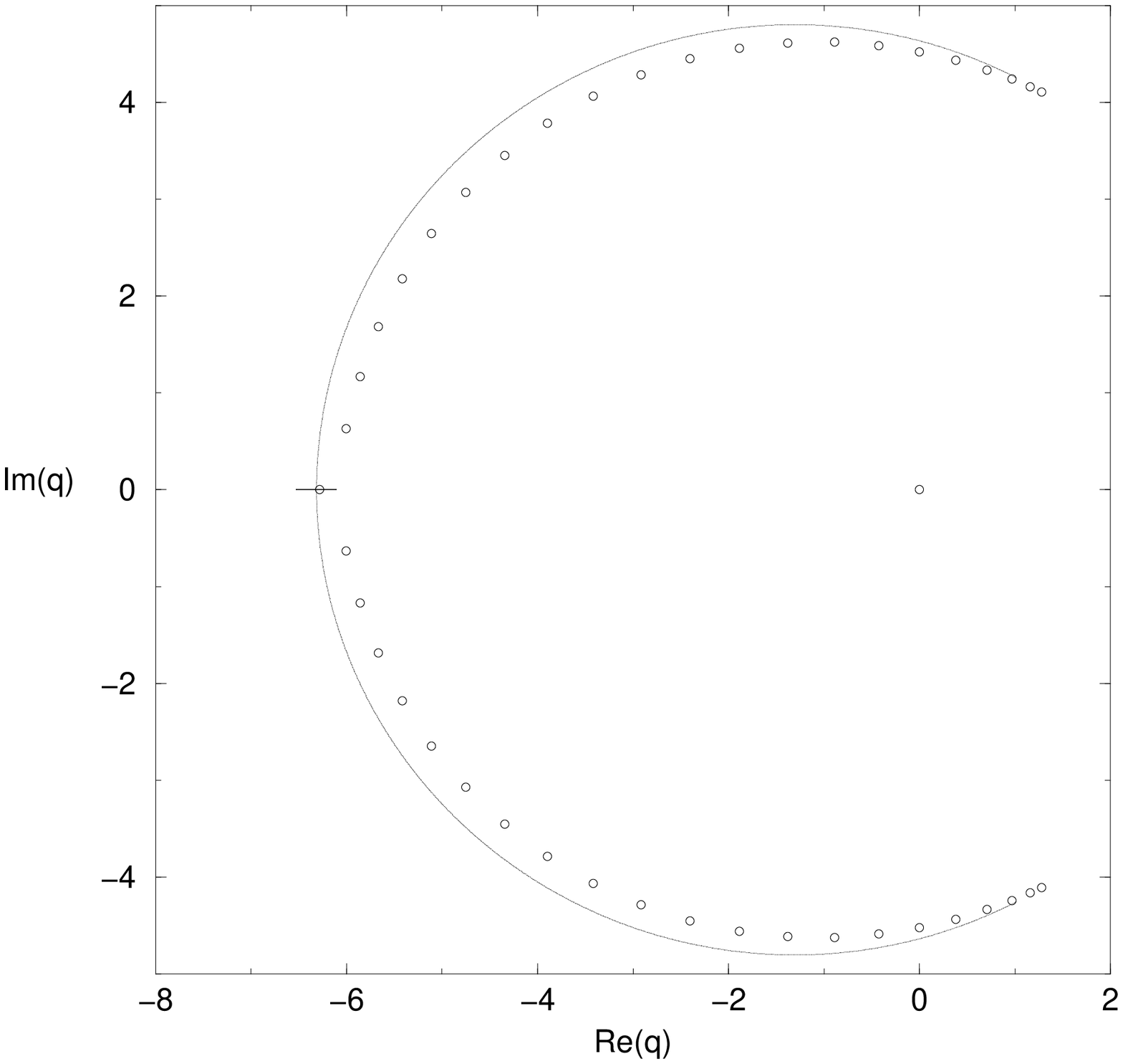}  
\end{center}
\vspace{-10mm}
\caption{\footnotesize{Locus ${\cal B}_q$: same as Fig. \ref{k4xy2a0p5}
for $a=2$ (i.e., $v=1$).}}
\label{k4xy2a2}
\end{figure}

We start with the value $a=0$ corresponding to the Potts antiferromagnet
at zero temperature.  In this case, $Z(S_m,q,v=-1)=P(S_m,q)$, where this
chromatic polynomial was given in eq. (\ref{psq}),
and the continuous locus ${\cal B}_q=\emptyset$. For $a$
in the finite-temperature antiferromagnetic range $0 < a < 0.0790075$,
${\cal B}_q$ consists of two self-conjugate arcs which cross the real $q$
axis between 2 and 3. For $0.0790075 < a < 0.704942$, 
${\cal B}_q$ consists of two arcs which are complex conjugates
of each other and do not cross the real $q$ axis. The endpoints of the
arcs occur at the branch points where the
function $R_{S12}$ in the square root is zero.
As $a$ increases through the value $0.704942$, the two arcs pinch
the real $q$ axis with $q$ smaller than 1, and ${\cal B}_q$ consists of a
single self-conjugate
arc crossing the positive real $q$ axis and a short line segment on the
$q$ axis. As $a$ reaches the infinite-temperature value 1, ${\cal B}_q$
shrinks to a point at the origin.  In the ferromagnetic range $a > 1$, the
self-conjugate arc crosses the negative real axis.  In Figs.
\ref{k4xy2a0p5}, \ref{k4xy2a0p9} and \ref{k4xy2a2} we show ${\cal B}_q$
and associated zeros of $Z$ in the $q$ plane for the antiferromagnetic
values $a=0.5$ and $a=0.9$, and the ferromagnetic value $a=2$.  Since the line
segment in Fig. \ref{k4xy2a0p9} is too small to be seen clearly, we note that
it extends outward from the point at 
which ${\cal B}_q$ crosses the real axis and covers the interval 
$0.378027 < q < 0.379970$.  The line segment is sufficiently long to be visible
in Fig. \ref{k4xy2a2}.  

\subsection{ ${\cal B}_u(\{S\})$ for fixed $q$}

\begin{figure}[hbtp]
\vspace{-20mm}
\centering
\leavevmode
\epsfxsize=2.5in
\begin{center}
\leavevmode
\epsffile{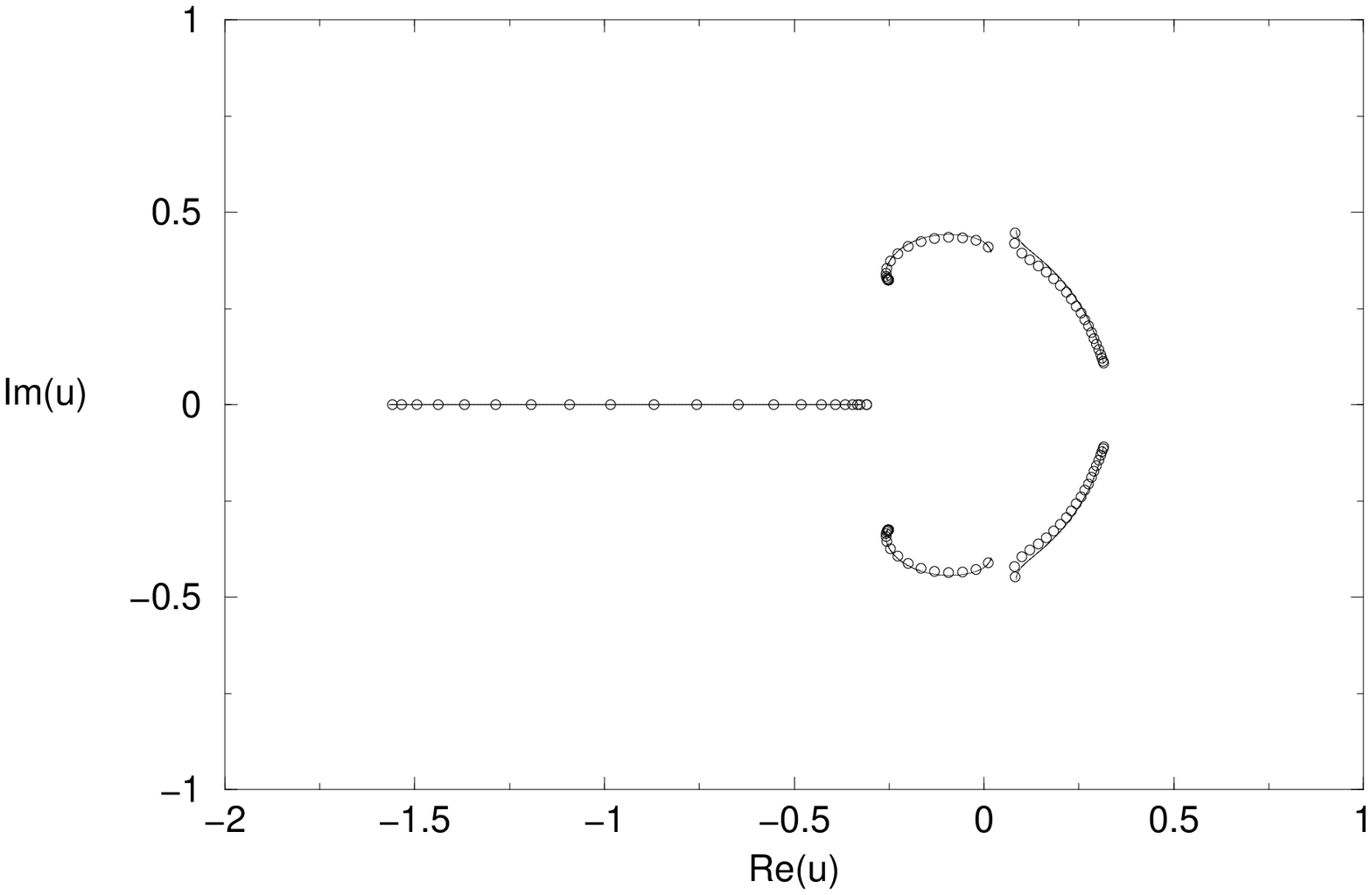}
\end{center}
\vspace{-20mm}
\caption{\footnotesize{Locus ${\cal B}_u$ for the $n \to \infty$ limit of
the $L_y=2$ $sq_d$ strip, with free longitudinal boundary conditions, 
$\{S\}$ with $q=10$. Zeros of $Z(S_m,q=10,v)$ in $u$ for $m=20$ are shown
for comparison.}}
\label{k4xy2q10}
\end{figure}

\begin{figure}[hbtp]  
\centering
\leavevmode
\epsfxsize=2.5in
\begin{center}
\leavevmode
\epsffile{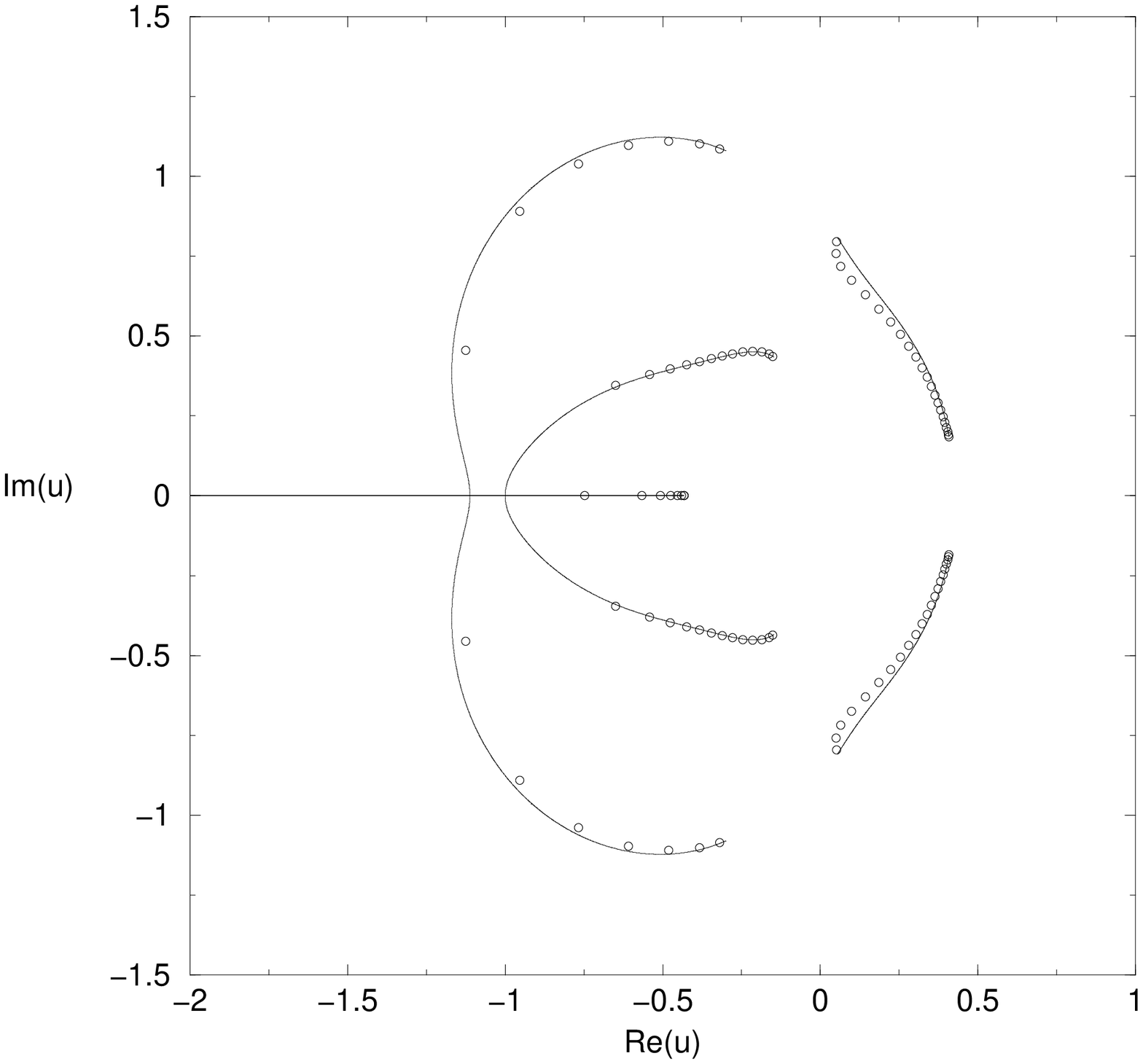} 
\end{center}
\vspace{-10mm}
\caption{\footnotesize{Locus ${\cal B}_u$: same as in Fig. \ref{k4xy2q10}
for $q=4$.}}
\label{k4xy2q4}
\end{figure}

\begin{figure}[hbtp]  
\centering
\leavevmode
\epsfxsize=2.5in
\begin{center}
\leavevmode
\epsffile{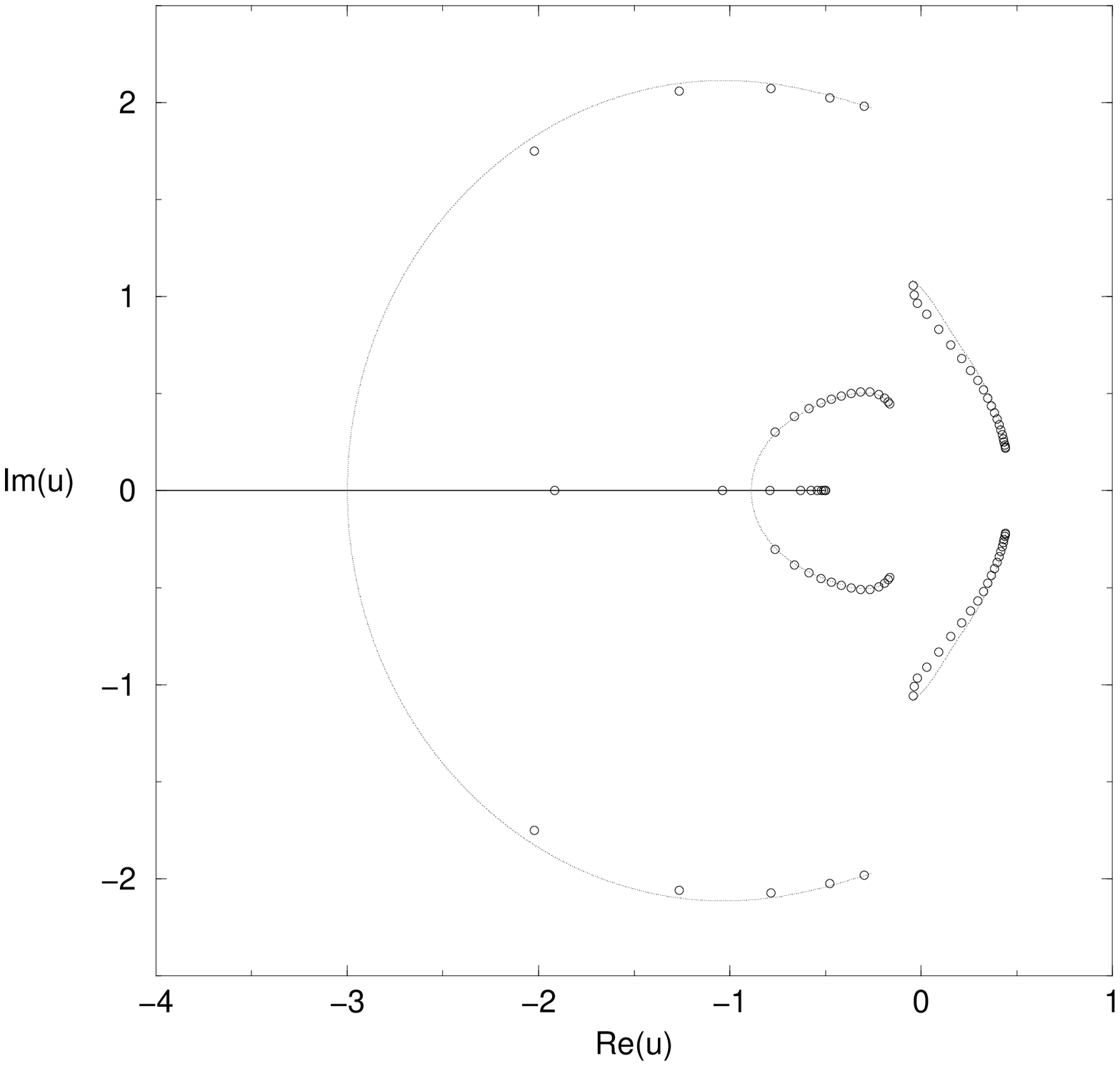} 
\end{center}
\vspace{-10mm}
\caption{\footnotesize{Locus ${\cal B}_u$: same as in Fig. \ref{k4xy2q10}
for $q=3$.}}
\label{k4xy2q3}
\end{figure}

\begin{figure}[hbtp]  
\centering
\leavevmode
\epsfxsize=2.5in
\begin{center}
\leavevmode
\epsffile{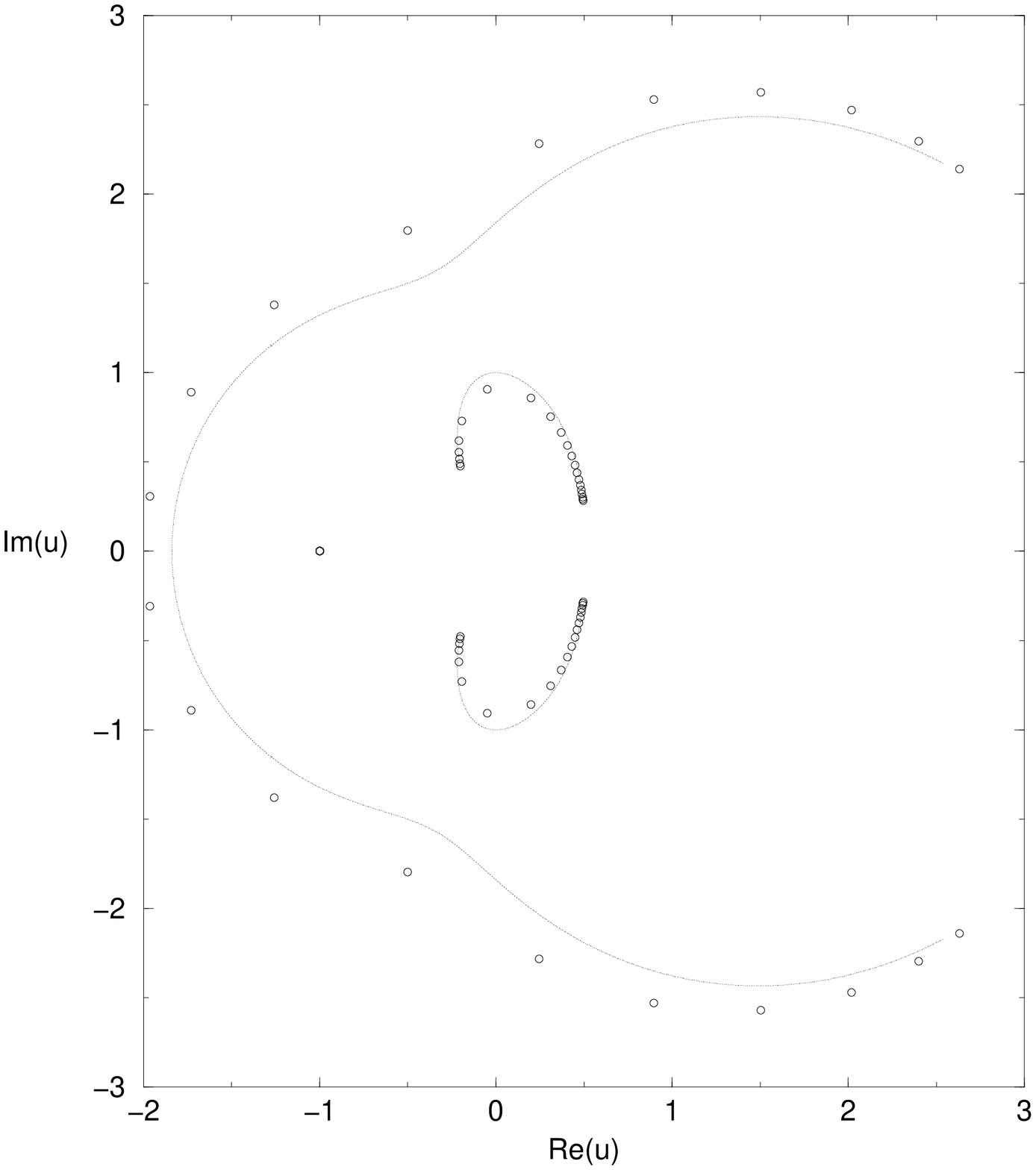} 
\end{center}
\vspace{-5mm}
\caption{\footnotesize{Locus ${\cal B}_u$: same as in Fig. \ref{k4xy2q10}
for $q=2$.}}
\label{k4xy2q2}
\end{figure}

\begin{figure}[hbtp]  
\centering
\leavevmode
\epsfxsize=2.5in
\begin{center}
\leavevmode
\epsffile{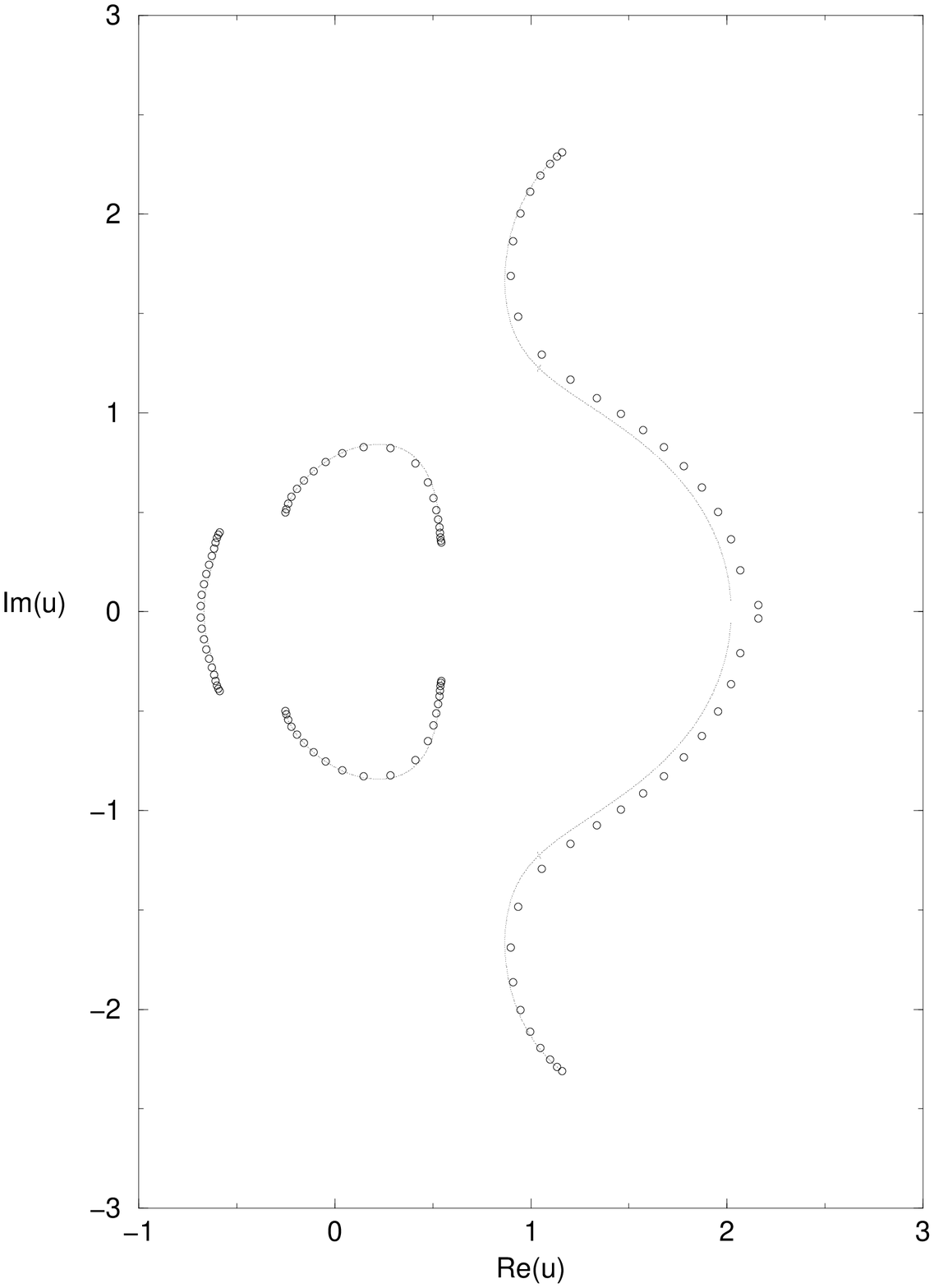} 
\end{center}
\caption{\footnotesize{Locus ${\cal B}_u$: same as in Fig. \ref{k4xy2q10}
for $q=1.5$.}}
\label{k4xy2q1p5}
\end{figure}

\begin{figure}[hbtp]  
\centering
\leavevmode
\epsfxsize=2.5in
\begin{center}
\leavevmode
\epsffile{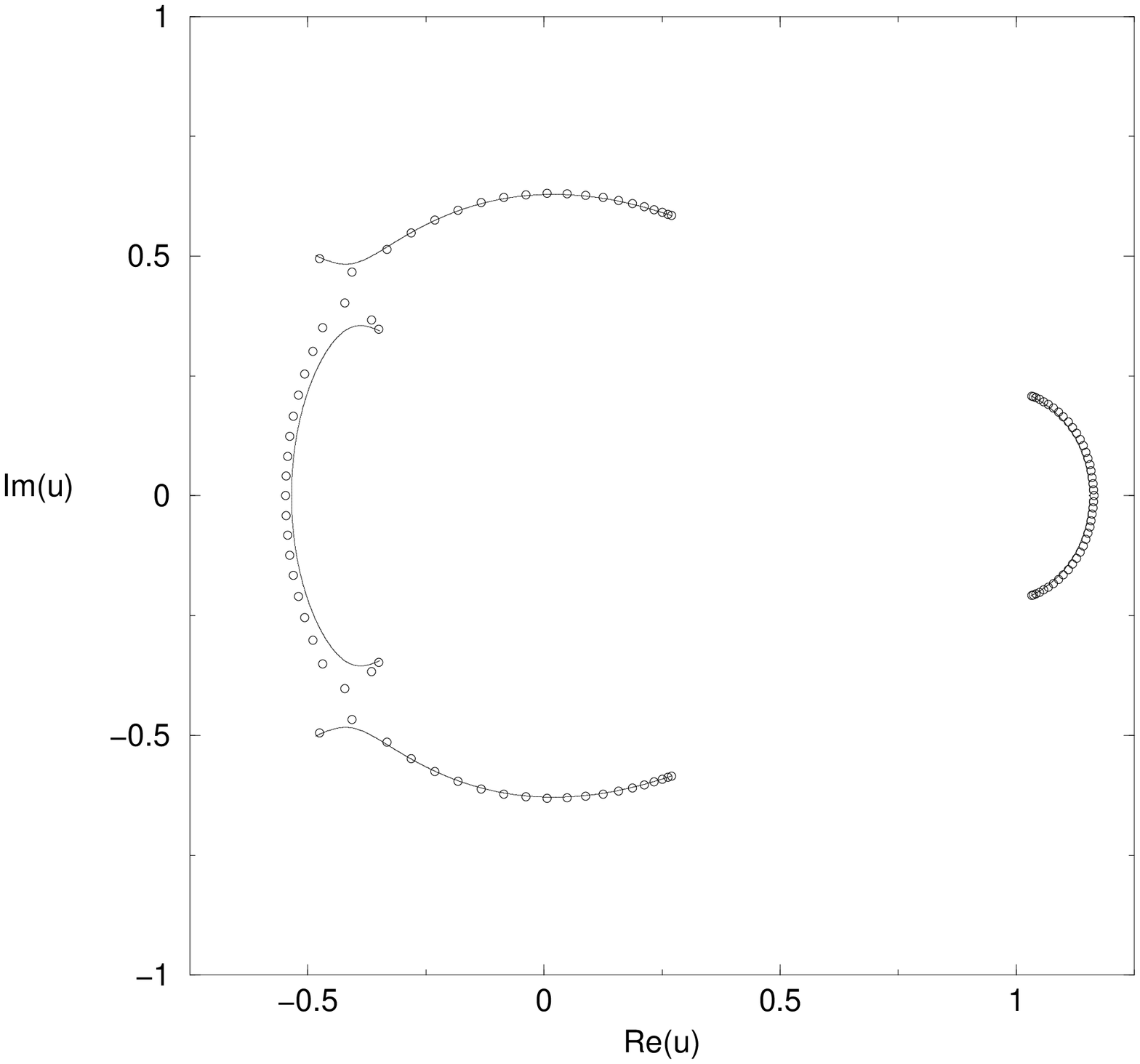} 
\end{center}
\vspace{-10mm}
\caption{\footnotesize{Locus ${\cal B}_u$: same as in Fig. \ref{k4xy2q10}
for $q=0.5$.}}
\label{k4xy2q0p5}
\end{figure}

We show several plots of the locus ${\cal B}_u$ for various values of $q$
in Figs. \ref{k4xy2q10} - \ref{k4xy2q0p5}.  In this section, for values of 
$q$ where
noncommutativity occurs, we display ${\cal B}_{qn}$.  Given the algebraic
structure of $\lambda_{S,j}$, $j=1,2$, the degeneracy of magnitudes  
$|\lambda_{S,1}|=|\lambda_{S,2}|$ and hence the locus ${\cal B}_u$ occurs   
where (i) $T_{S12}=0$, (ii) $R_{S12}=0$, (iii) if $q$ is real, where
$R_{S12} < 0$ so that the square root is pure imaginary, and (iv)
elsewhere for complex $u$, where the degeneracy condition is satisfied.
The locus ${\cal B}_u$ does not enclose regions, and $\lambda_{S,1}$ is
dominant everywhere in the $u$ plane and degenerate in magnitude with
$\lambda_{S,2}$ on ${\cal B}_u$.

For large values of $q$, we find that ${\cal B}_u$ consists of two pairs of
complex conjugate arcs, and a line segment on the negative $u$ axis. The ten
endpoints are the branch point zeros of $\sqrt{R_{S12}}$, and the line segment
is the solution to the condition (iii) above.  As $q$ decreases, the locus
${\cal B}$ changes to consist of two self-conjugate arcs, as shown in
Figs. \ref{k4xy2q4} and \ref{k4xy2q3}.  For $2 < q < 3$, one self-conjugate arc
crosses the positive $u$ axis at the point where the condition (i) is
satisfied. At $q=2$, there is only one pair of complex conjugate arcs and a
self-conjugate arc, the latter of which crosses 
the negative $u$ axis. For $1 < q < 2$, ${\cal B}_u$
consists of five disjoint arcs, as illustrated for the value $q=3/2$ in
Fig. \ref{k4xy2q1p5}. At $q=1$, $({\cal B}_u)_{qn}$ is an oval (the solution to
the degeneracy equation $2|u^3(u-1)^2|=1$) that crosses the real axis at 
approximately 1.418 and $-0.584$. 

For $0 < q < 1$, ${\cal B}$ consists of two self-conjugate arcs, one pair
of complex conjugate arcs, and a short line segment on the real $u$ axis 
as illustrated for the value $q=1/2$ in Fig. \ref{k4xy2q0p5} (the line segment
is too short to be clearly visible in this figure; it covers the interval 
$1.15474 < q < 1.15598$). For $q=0$,
the locus $({\cal B}_u)_{qn}$ consists of two pairs of complex conjugate
arcs.

\section{Cyclic and M\"obius Strips of the $sq_d$ Lattice} 

\subsection{Results for $Z$}

By using either an iterative application of the deletion-contraction theorem
for Tutte polynomials and converting the result to $Z$, or a transfer matrix 
method, one can calculate the
partition function for the cyclic strip graphs of arbitrary
length, $Z(G,q,v)$, $G=L_m$.  We have used both methods as checks on
the calculation.  Our results
have the general form (\ref{zgsum}) with $N_\lambda=5$ and are
\beq
\lambda_{L,1}=2v^2
\label{lam1}
\eeq
\beq
\lambda_{L,(2,3)}=\frac{v}{2}\biggl [ T_{23} \pm \sqrt{R_{23}} \ \biggr ]
\label{lam23}
\eeq
where
\beq
T_{23}=v^4+5v^3+10v^2+12v+2q
\label{t23}
\eeq
\beq
R_{23}=144v^2+32qv+4q^2+v^8+10v^7+45v^6+196v^4+116v^5-12qv^3-4qv^4+224v^3
\label{r23}
\eeq
and
\beq
\lambda_{L,4}=\lambda_{S,1} \ , \quad \lambda_{L,5}=\lambda_{S,2} 
\label{lam45}
\eeq
where $\lambda_{S,j}$, $j=1,2$, were given above in (\ref{lams}). 

The coefficients are 
\beq
c_{L,1}=\frac{q(q-3)}{2}
\label{c1}
\eeq
\beq
c_{L,2}=c_{L,3}=q-1
\label{c23}
\eeq
\beq
c_{L,4}=c_{L,5}=1
\label{c45}
\eeq
These can be expressed in terms of Chebyshev polynomials \cite{cf}:
\beq
c^{(d)}=U_{2d}\Bigl (\frac{\sqrt{q}}{2} \Bigr )
\label{cd}
\eeq
where $U_n(x)$ is the Chebyshev polynomial of the second kind, defined by 
\beq
U_n(x) = \sum_{j=0}^{[\frac{n}{2}]} (-1)^j {n-j \choose j} (2x)^{n-2j}
\label{undef}
\eeq
where the notation $[\frac{n}{2}]$ in the upper limit on the summand means 
the integral part of $\frac{n}{2}$.  The first few of the $c^{(d)}$'s are 
$c^{(0)}=1$, $c^{(1)}=q-1$, and $c^{(2)}=q^2-3q+1$.  Thus, for the present case
we have 
\beq
c_{L,1}=\frac{1}{2}(c^{(2)}-c^{(0)})
\label{cl1cheb}
\eeq
$c_{L,2}=c_{L,3}=c^{(1)}$, and $c_{L,4}=c_{L,5}=c^{(0)}$.
The partition function $Z(L_m,q,v)$ satisfies the general relations 
(\ref{zq0})-(\ref{za1}) and (\ref{pq123}).  

Our main interest here is in large $m$ and the $m \to \infty$ limit.  However,
for completeness, we make the following remark.  If $m \ge 3$, then $L_m$ is a
(proper) graph, but the $m=1$ and $m=2$ cases requires special consideration;
in these cases, $L_m$ degenerates and is not a proper
graph\footnote{\footnotesize{A proper graph has no multiple edges or loops,
where a loop is an edge that connects a vertex to itself.  A multigraph may
contain multiple edges, but no loops, while a pseudograph may contain both
multiple edges and loops \cite{bbook,boll}.}}.  $L_2$ is the multigraph
obtained from the complete graph $K_4$ by doubling all edges except two edges
that do not have a common vertex.  $L_1$ is the pseudograph obtained by
connecting two vertices with three edges and adding a loop to each vertex.  Our
calculation of $Z(L_m,q,v)$ and the corresponding Tutte polynomial $T(L_m,x,y)$
applies not just for the uniform cases $m \ge 3$ but also for the special cases
$m=1,2$ if for $m=2$ one includes the multiple edges and for $m=1$ the multiple
edges and loops in the evaluation of (\ref{zfun}), (\ref{ham}), and
(\ref{cluster}).  Note that in the $T=0$ case for the antiferromagnet, the
resulting partition function, or equivalently, the chromatic polynomial, is not
sensitive to multiple edges, i.e. is the same for a graph in which two vertices
are connected by one edge or multiple edges; however, the general partition
function (Tutte polynomial) is sensitive to multiple edges.  The chromatic
polynomial is sensitive to loops and vanishes identically when a pseudograph
has any loops.

\subsection{Special values and expansions of $\lambda$'s}

We discuss some special cases here. First, for the zero-temperature Potts
antiferromagnet, i.e. the case $a=0$ ($v=-1$), the partition function
$Z(L_m,q,v)$ reduces, in accordance with the general result
(\ref{zp}), to the chromatic polynomial $P(L_m,q)$
calculated in \cite{readcarib88}.  In this special case 
\beq
\lambda_{L,1}=2
\label{lam1chrom}
\eeq
\beq
\lambda_{L,2}=2(3-q)
\label{lam2chrom}
\eeq
\beq
\lambda_{L,3}=0
\label{lam3chrom}
\eeq
\beq
\lambda_{L,4}=(q-2)(q-3)
\label{lam4chrom}
\eeq
\beq
\lambda_{L,5}=0
\label{lam5chrom}
\eeq
(where these presume certain choices of branch cuts for square roots, so that,
e.g., $\sqrt{(3-q)^2}=3-q$, etc.). 

For the infinite-temperature value $a=1$, we have $\lambda_{L,j}=0$ for
$j=1,2,3,5$, while $\lambda_{L,4}=q^2$, so that $Z(L_m,q,a=1)= q^{2m} =
q^n$, in accord with the general result (\ref{za1}).

At $q=0$, we find (with appropriate choices of branch cuts)
\beq
\lambda_{L,1}=2v^2
\label{lam1q0}
\eeq
\beq
\lambda_{L,2}=\lambda_{L,4}=\frac{1}{2}\Bigl [ (v+3)(v^2+2v+4) +
\sqrt{144+v^6+10v^5+45v^4+196v^2+116v^3+224v} \ \Bigr ]
\label{lam23q0}
\eeq
\beq
\lambda_{L,3}=\lambda_{L,5}=\frac{1}{2}\Bigl [ (v+3)(v^2+2v+4) - 
\sqrt{144+v^6+10v^5+45v^4+196v^2+116v^3+224v} \ \Bigr ]
\label{lam45q0}
\eeq
Since there are two degenerate dominant terms, namely
$\lambda_2=\lambda_4$, it follows that 
\beq
q=0 \quad {\rm is \ on} \quad {\cal B}_q(\{L\}) \quad \forall \ a \ .
\label{q0onb}
\eeq
This was also true of the circuit graph and cyclic and M\"obius square and
triangular strips
with $L_y=2$ for which the general Potts model partition function (Tutte
polynomial) was calculated in \cite{a}.  For $q=0$, the coefficients
$c_{L_1}=0$, $c_{L_2}=c_{L_3}=-1$, and $c_{L,4}=c_{L,5}=1$ so that the equal
terms cancel each other pairwise, yielding $Z(L_m,q=0,v)=0$, in accordance
with the general result (\ref{zq0}).  The noncommutativity
(\ref{fnoncomm}) occurs here: $\exp(f_{nq})=0$, while
$|\exp(f_{qn})|=|\lambda_{L,4}|^{1/2}$. 

At $q=1,2$ we again encounter noncommutativity in the calculation of
the free energy: for $q=1$, eq. (\ref{zq1}) yields the result
$Z(L_m,q=1,a)
=a^{5m}$, whence
\beq
\exp(f_{nq}(\{L\},q=1,a))=a^{5/2}
\label{fnqq1}
\eeq
while $f_{qn}$ depends on which phase one is in for a given value of $a$.

To discuss the special case $q=2$, we first observe that 
\beq
\lim_{q \to 2} \lim_{v \to -1} \lambda_{L,3} \ne 
\lim_{v \to -1} \lim_{q \to 2} \lambda_{L,3}
\label{lam3noncomq2}
\eeq
Specifically, 
\beq
\lambda_{L,3}(v=-1) = 0 \quad {\rm if} \ \ q \ne 2
\label{lam3a0}
\eeq
so that the left-hand side of (\ref{lam3noncomq2}) is zero, while
\beq
\lambda_{L,3}(q=2) = 2v^2 \quad {\rm if} \ \ v \ne -1
\label{lam3q2}
\eeq
so that the right-hand side of (\ref{lam3noncomq2}) is 2. Similarly,
$\lambda_{L,5}(v=-1) = 0$ if $q \ne 2$, but $\lambda_{L,5}(q=2)$ is
nonzero if $v \ne -1$ as given below.
Thus, for the special case $q=2$, with the understanding that in cases where
the noncommutativity (\ref{lam3noncomq2}) holds, we take the limit $q=2$ first
for general $v$, with $v \ne -1$, we have
\beq
\lambda_{L,1}=\lambda_{L,3}=2v^2
\label{lam1q2}
\eeq
\beq
\lambda_{L,2}=v(v+1)(v+2)(v^2+2v+2)
\label{lam2q2}
\eeq
\beq
\lambda_{L,(4,5)}=\frac{(v+1)(v+2)}{2}\biggl [ v^3+2v^2+2v+2 \pm 
\sqrt{v^6+4v^5+8v^4+4v^3+4v^2+8v+4} \biggr ] 
\label{lam45q2}
\eeq
Further, for $q=2$, $c_{L,1}=-1$, while $c_{L,j}=1$ for $2 \le j \le 5$; hence,
the $(\lambda_{L,1})^m$ and $(\lambda_{L,3})^m$ terms cancel each other and
make no contribution to $Z$, which reduces to
\beq
Z(L_m,q=2,v)=\sum_{j=2,4,5} (\lambda_{L,j})^m
\label{zq2}
\eeq
Hence also, $f_{qn} \ne f_{nq}$ at $q=2$.

In order to study the zero-temperature critical point in the ferromagnetic
case and also the properties of the complex-temperature phase diagram, we
calculate the $\lambda_{L,j,u}$'s corresponding to the $\lambda_{L,j}$'s,
using eq. (\ref{lamu}).  In the vicinity of the point $u=0$ we have
\beq
\lambda_{L,1,u}=  2u^3(1-u)^2
\label{lam1rtaylor}
\eeq
and the Taylor series expansions 
\beq
\lambda_{L,2,u}=1-u^4+2(q-2)u^5 + O(u^7)
\label{lam2rtaylor}
\eeq
\beq
\lambda_{L,3,u}=2u^3+2(q-4)u^5 + O(u^7)
\label{lam3rtaylor}
\eeq
\beq
\lambda_{L,4,u}=1+(q-1)u^4+4(q-1)u^5+O(u^7)
\label{lam4rtaylor}
\eeq
\beq
\lambda_{L,5,u}=2u^3+4(q-2)u^4 + O(u^5) \ . 
\label{lam5rtaylor}
\eeq 
Therefore, at $u=0$, $\lambda_{L,2,u}$ and $\lambda_{L,4,u}$ are dominant
and the boundary ${\cal B}_u$ is determined as the solution to the degeneracy
of magnitudes of these dominant $\lambda_{L,j,u}$'s, i.e.,
$|\lambda_{L,2,u}|=|\lambda_{L,4,u}|$, so that the point $u=0$ is on ${\cal
B}_u$ for any $q$, with the understanding that for the values of $q$ where the
noncommutativity (\ref{fnoncomm}) and (\ref{bnoncomm}) occurs, we are referring
to ${\cal B}_{qn}$ rather than ${\cal B}_{nq}$.

To determine the angles at which the branches of ${\cal B}_u$ cross
each other at $u=0$, we write $u$ in polar coordinates as $u=re^{i\theta}$,
expand the degeneracy equation $|\lambda_{L,2,u}|=|\lambda_{L,4,u}|$, for small
$r$, and obtain $qr^4\cos(4\theta)=0$, which implies that (for $q \ne 0$) in
the limit as $r=|u| \to 0$,
\beq
\theta = \frac{(2j+1)\pi}{8} \ , \quad j=0,1,...,7
\label{thetau}
\eeq
or equivalently, $\theta=\pm \pi/8$, $\pm 3\pi/8$, $\pm 5\pi/8$, and 
$\pm 7\pi/8$.  Hence there are eight curves forming four branches on 
${\cal B}_u$ 
intersecting at $u=0$, with an angle of $\pi/4$ between each adjacent pair of
curves at $u=0$. The point $u=0$ is thus a multiple point on
the algebraic curve ${\cal B}_u$, in the technical terminology of algebraic
geometry (i.e., a point where several branches of an algebraic curve cross
\cite{alg}).  In the vicinity of the origin, $u=0$, these branches define eight
corresponding complex-temperature phases: the paramagnetic (PM) phase for 
$-\pi/8 \le \theta \le \pi/8$, together with seven $O_j$ phases extending 
outward in the wedges $(2j-1)\pi/8 \le \theta \le (2j+1)\pi/8$ for
$j=1,..,7$. These 
form one self-conjugate phase, $O_4=O_4^*$, and the complex conjugate pairs 
of phases $O_1=O_7^*$, $O_2=O_6^*$, and $O_3=O_5^*$.

For $q=2$ and for $q=4$ the Potts antiferromagnet on the infinite-length, width
$L_y=2$ strip of the $sq_d$ lattice has a zero-temperature critical point. In
the $q=2$ Ising case, this involves frustration. In order to study the $T=0$
critical point for the $L_y=2$ strip for these two values of $q$, it is useful
to calculate expansions of the $\lambda_{L,j}$'s. Only $\lambda_{L,4}$ and
$\lambda_{L,2}$ are necessary for physical thermodynamic properties, while the
full set of $\lambda_{L,j}$, $j=1,2,..,5$ is, in general, necessary for the
study of the singular locus ${\cal B}$.

For $q=2$, besides the exact expressions
$\lambda_{L,1}=\lambda_{L,3}=2(a-1)^2$ and 
$\lambda_{L,2}=-a+a^5$, we have the expansions
\beq
\lambda_{L,4} = a + 4a^4 + 9a^5 + O(a^6)
\label{lam4q2taylor}
\eeq
\beq
\lambda_{L,5} = 2a^2 - 4a^4 - 8a^5 + O(a^6)
\label{lam5q2taylor}
\eeq
As shown above, for $f_{nq}$ and ${\cal B}_{nq}$, where one sets $q=2$
first and then takes $n \to \infty$, $\lambda_{L,j}$, $j=1,3$, make no
contribution, and  ${\cal B}_{nq}$ is determined by the degeneracy in
magnitude of the dominant terms among $\lambda_{L,j}$, $j=2,4,5$.  
From the expansions
(\ref{lam4q2taylor}) and (\ref{lam5q2taylor}), it follows that in the
neighborhood of the point $a=0$, $({\cal B}_a)_{nq}$ is determined by the
equation $|\lambda_{L,2}|=|\lambda_{L,4}|$.  Writing $a=\rho e^{i\phi}$
and expanding the degeneracy equation $|\lambda_{L,2}|=|\lambda_{L,4}|$
for small $r$, we obtain $8\rho^5\cos(3\phi)=0$, which implies that in the
limit as $r=|a| \to 0$,
\beq
\phi = \frac{(2j+1)\pi}{6} \ , \quad j=0,1,...,5
\label{phia}
\eeq
or equivalently, $\phi=\pm \pi/6$, $\phi=\pm \pi/2$, and $\phi=\pm 5\pi/6$.   
Hence there are six curves forming three branches of ${\cal B}_a$ 
intersecting at $a=0$ and the angles between successive branches as they cross
at this point are $\pi/3$. 
The point $a=0$ is thus a multiple point on ${\cal B}_u$. In 
the vicinity of the origin, $a=0$, these branches define six 
complex-temperature phases: the paramagnetic (PM) phase for
$-\pi/6 \le \theta \le \pi/6$, together with the phases $O_j$ for $1 \le j
\le 5$, with $O_j$ occupying the sector $(2j-1)\pi/6 \le \theta \le
(2j+1)\pi/6$. Note that $O_3=O_3^*$, $O_4=O_2^*$, and $O_5=O_1^*$. 
$\lambda_{L,4}$ is dominant in the PM phase and in the $O_2$ and $O_2^*$
phases, while $\lambda_{L,2}$ is dominant in the $O_1$, $O_1^*$, and $O_3$
phases.

For $q=4$, besides the exact expression $\lambda_{L,1}=2(a-1)^2$,   
we calculate the expansions
\beq
\lambda_{L,2} = -2 - 2a + 3a^2 + O(a^3) 
\label{lam2q4taylor}
\eeq
\beq
\lambda_{L,3} = a - a^2 + O(a^3) 
\label{lam3q4taylor}
\eeq
\beq
\lambda_{L,4} = 2 + 14a - 31a^2 + O(a^3)
\label{lam4q4taylor}
\eeq
\beq
\lambda_{L,5} = -3a + 33a^2 + O(a^3) \ . 
\label{lam5q4taylor}
\eeq
It follows that in the neighborhood of the point $a=0$, $({\cal
B}_a)_{nq}$ is determined by the equation
$|\lambda_{L,1}|=|\lambda_{L,4}|$ and passes through $a=0$ vertically on
the imaginary axis.

\subsection{${\cal B}_q(\{L\})$ for fixed $a$}

\subsubsection{Antiferromagnetic case $0 \le a \le 1$}

For $a=0$, i.e., the $T=0$ limit of the Potts antiferromagnet, the locus 
${\cal B}_q$ consists of the union of the two circles 
\beq
{\cal B}_q : \ |q-2|=2 \ \cup \ |q-3|=1 \quad {\rm for} \quad a=0
\label{bfora0}
\eeq
so that there ${\cal B}_q$ crosses the real $q$ axis at $q=0,2,4$, so that 
$q_c=4$; furthermore, the point $q=4$ is a multiple point on ${\cal B}_q$, 
where the smaller and larger circles intersect (shown in Fig. 4 in \cite{k}).  
This multiple point is a tacnode, i.e. the curves that intersect at this 
point have the same (vertical) slopes.  In region $R_1$ forming the exterior
of the larger circle, $|q-2| > 2$, $\lambda_{L,4}$ is dominant.  In region 
$R_2$ forming the interior of the smaller circle, $|q-3| < 1$, $\lambda_{L,1}$ 
is dominant, while in region $R_3$ comprised of the crescent-shaped area 
inside the larger circle and outside the smaller circle, i.e., with 
$|q-2| < 2$ and $|q-3| > 1$, $\lambda_{L,2}$ is dominant. 

\begin{figure}[hbtp]
\centering
\leavevmode
\epsfxsize=2.5in
\begin{center}
\leavevmode
\epsffile{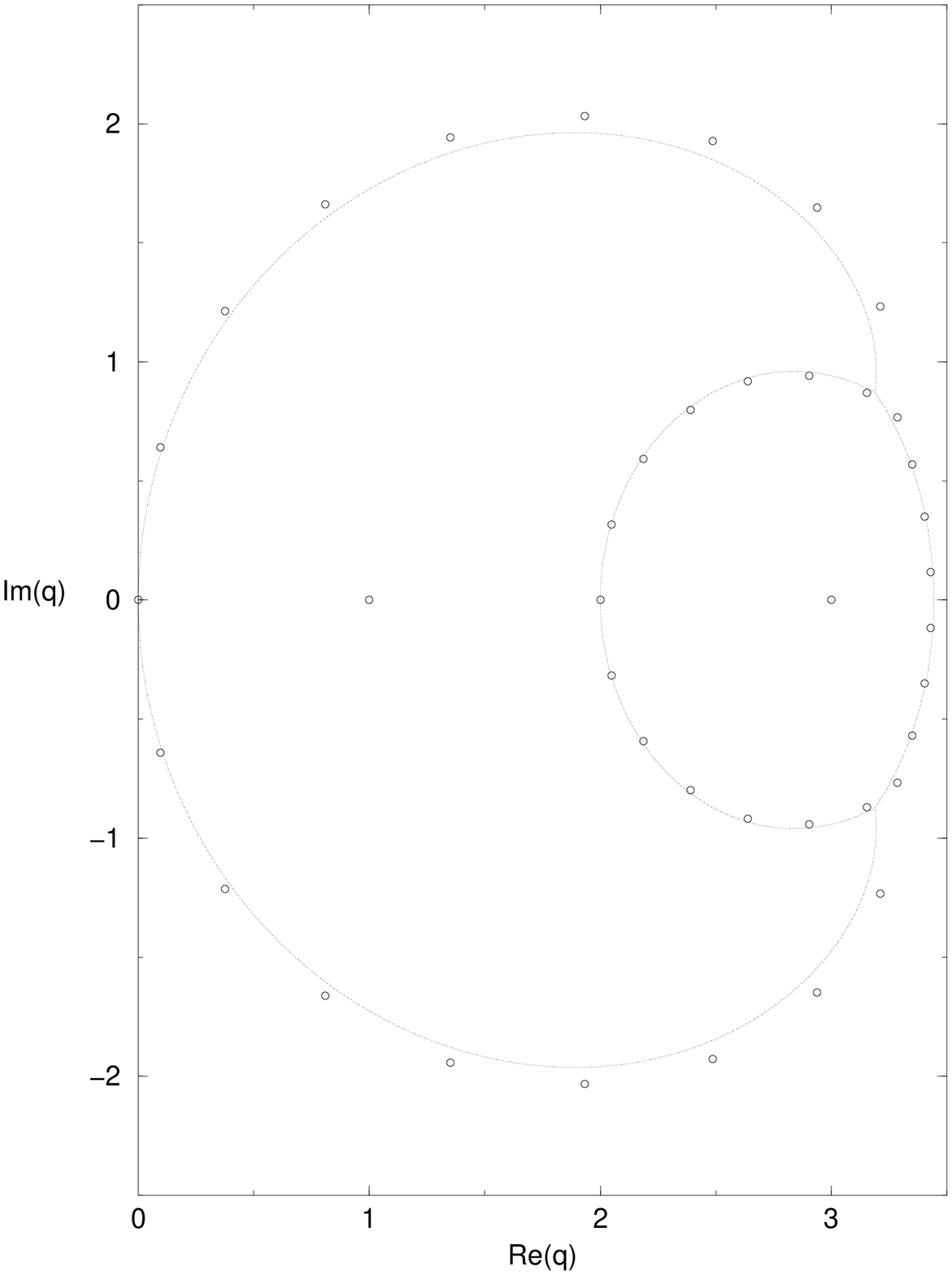}
\end{center}
\caption{\footnotesize{Locus ${\cal B}_q(\{L\})$ for the $n \to \infty$
limit of the $L_y=2$ $sq_d$ strip for $a=0.1$.}}
\label{k4pxy2a0p1}
\end{figure}

\begin{figure}[hbtp]
\centering
\leavevmode
\epsfxsize=2.5in
\begin{center}
\leavevmode
\epsffile{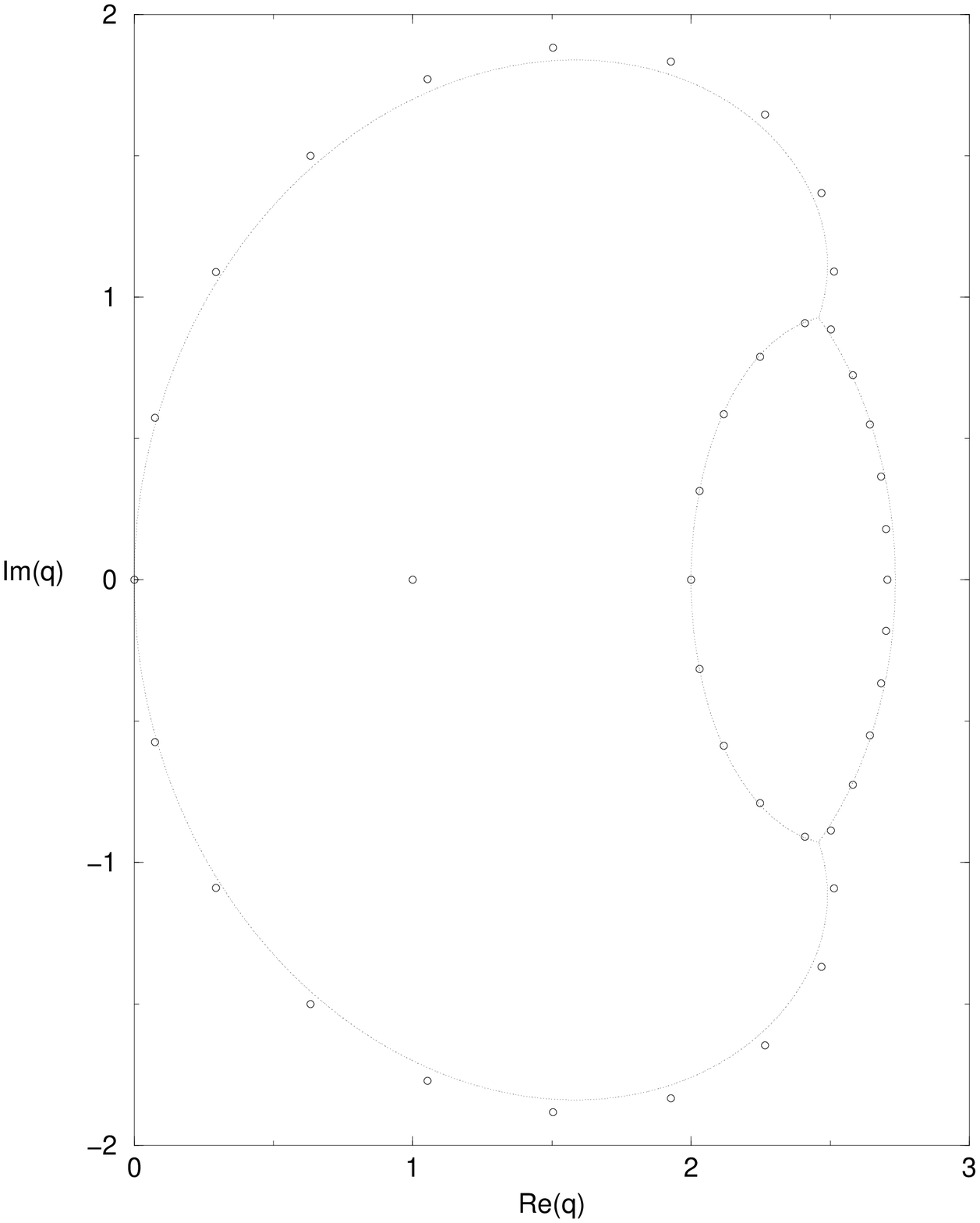}
\end{center}
\caption{\footnotesize{Locus ${\cal B}_q(\{L\})$: same as Fig.
\ref{k4pxy2a0p1} for $a=0.25$.}}
\label{k4pxy2a0p25}
\end{figure}

\begin{figure}[hbtp]
\centering
\leavevmode
\epsfxsize=2.5in
\begin{center}
\leavevmode
\epsffile{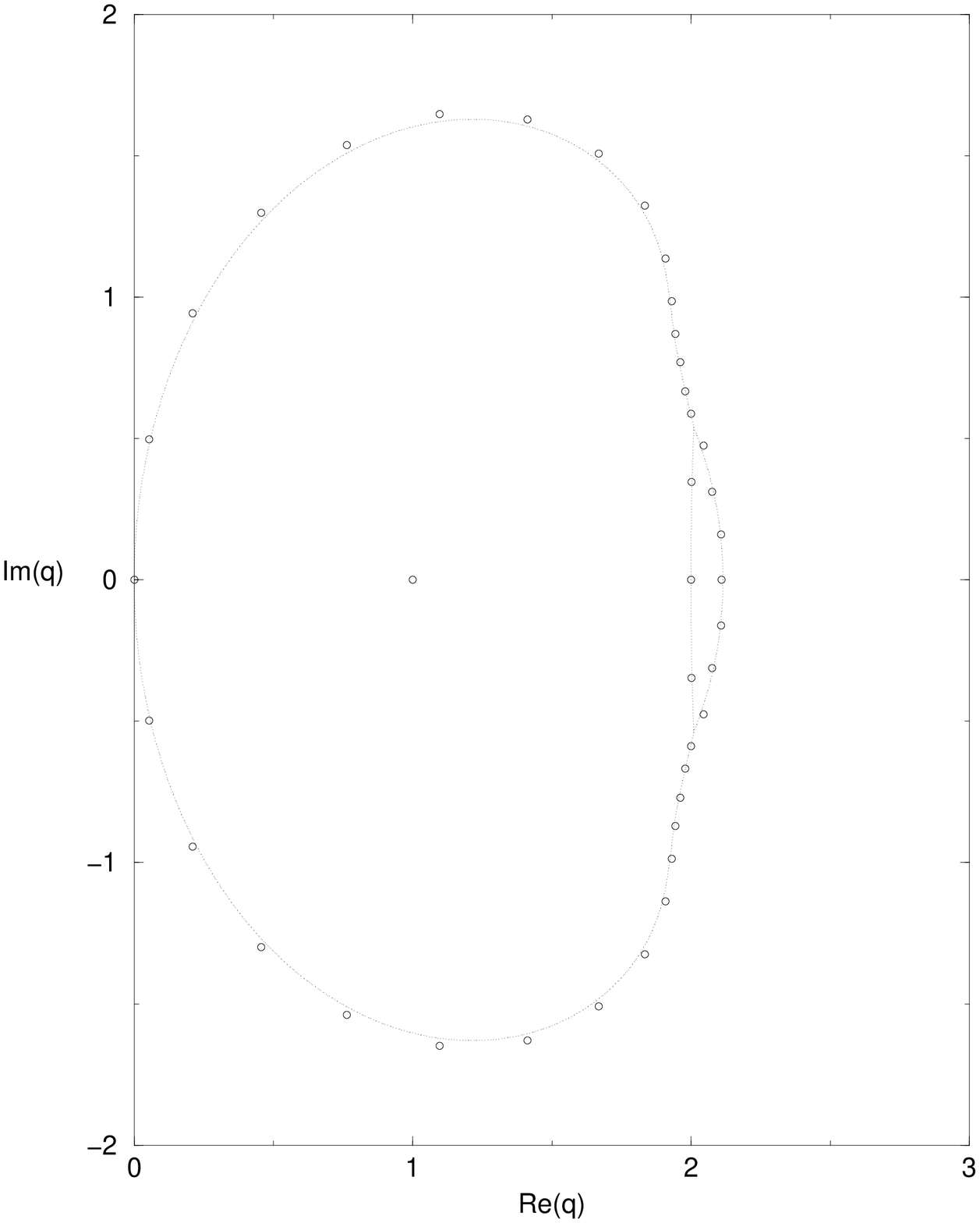}
\end{center}
\caption{\footnotesize{Locus ${\cal B}_q(\{L\})$: same as Fig.
\ref{k4pxy2a0p1} for $a=0.4$.}}
\label{k4pxy2a0p4}
\end{figure}

\begin{figure}[hbtp]
\centering
\leavevmode
\epsfxsize=2.5in
\begin{center}
\leavevmode
\epsffile{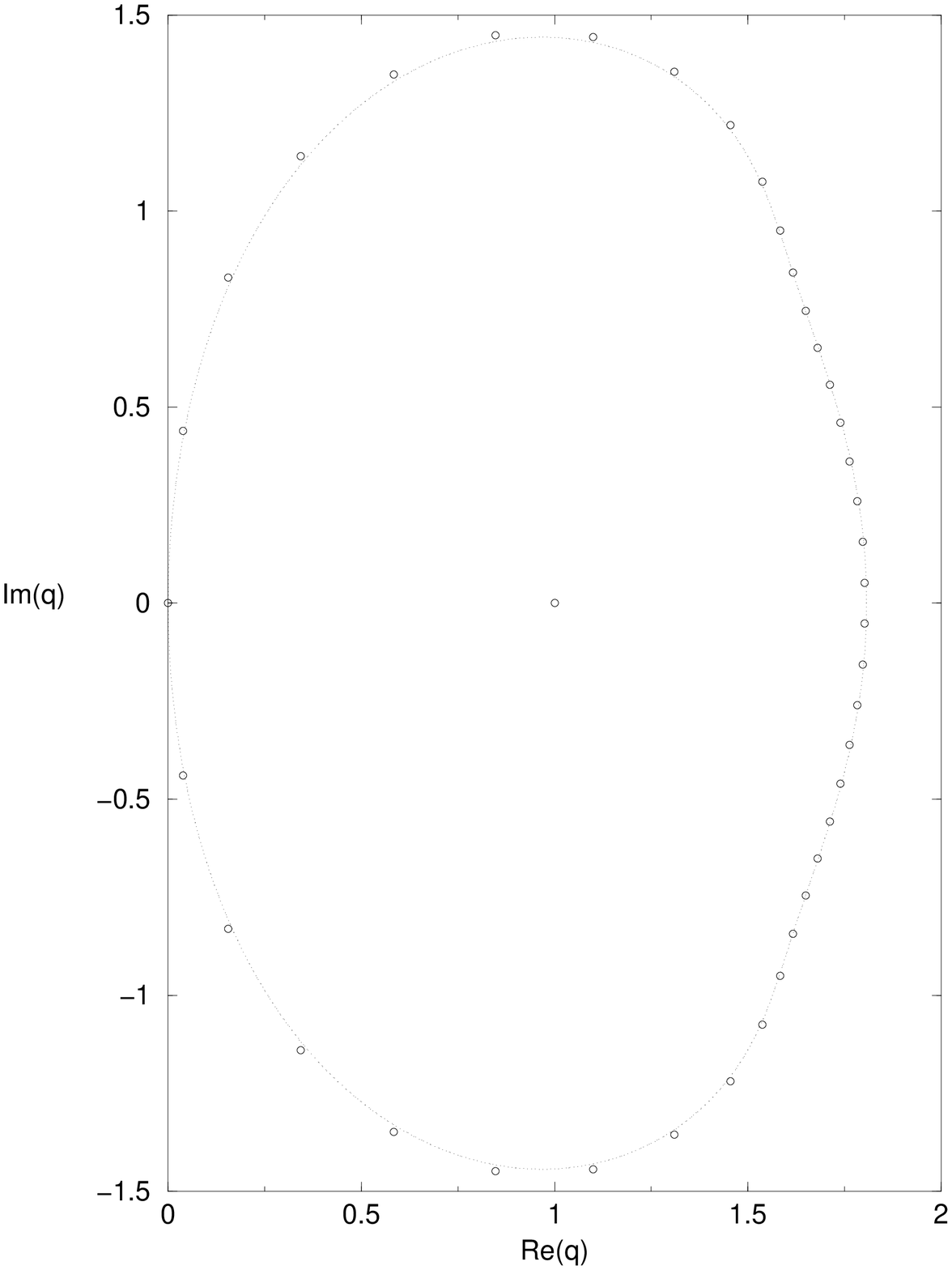}
\end{center}
\caption{\footnotesize{Locus ${\cal B}_q(\{L\})$: same as Fig.
\ref{k4pxy2a0p1} for $a=0.5$.}}
\label{k4pxy2a0p5}
\end{figure}

As the temperature increases from 0 to infinity for the antiferromagnet, i.e.,
as $a$ increases from 0 to 1, ${\cal B}_q$ contracts in to the origin, $q=0$. 
Using our exact calculation of the Potts partition function for arbitrary $T$, 
we have determined ${\cal B}_q$ for general $a$.  In Figs.
\ref{k4pxy2a0p1}-\ref{k4pxy2a0p5} we show some illustrative plots for the
Potts antiferromagnet.  As $a$ increases from 0, the tacnodal multiple point
that existed on ${\cal B}_q$ for $a=0$ disappears and the locus has the
appearance illustrated by Fig. \ref{k4pxy2a0p1}. This is qualitatively similar
to what we found for ${\cal B}_q$ for the ($L_x \to \infty$ limit of the) 
$L_y=2$ strip of the triangular lattice \cite{ta} as $a$ increased from 0. 
The middle crossing point at $q=2$ remains fixed, independent of 
$a$, for $a$ in the range $0 < a < a_m$, where 
\beq
a_m \simeq 0.4294445
\label{am}
\eeq
is the real solution of the equation
\beq
2a^3+3a^2+3a-2=0
\label{meq}
\eeq
The reason for this 
is that this point is determined by the degeneracy equation $|\lambda_{L,1}|=
|\lambda_{L,2}|$ where these are leading terms, and they cease to be
leading
at $q=2$ as the right boundary sweeps leftward past this point; in turn, this
occurs as $q_c=2$, which yields the equation (\ref{meq}).  As $a$ increases
from 0 to $a_m$, region $R_2$ contracts in size, and finally disappears
altogether as $a$ increases through the value $a_m$. For $a > a_m$, 
$\lambda_{L,4}$ is dominant in the neighborhood of $q=2$ as well as for 
$q > 2$.  Again, this feature, that the crossing at $q=2$ remains fixed as $a$
increases until the right-most part of ${\cal B}$ sweeps past it, thereby
removing the region $R_2$, is qualitatively the same as what we found for the 
$L_y=2$ strip of the triangular lattice \cite{ta}.  
The value of $a$ at which $R_2$ disappears in that case was 
$(-3+\sqrt{17})/4=0.280776...$.  Note that ${\cal B}$ for the $L_y=2$ square
lattice strip is different in that in the $a=0$ case, $q_c=2$ and there are
only two regions that contain intervals on the real axis: $R_1$ for $q < 0$ and
$q > q_c=2$, and $R_2$ for $0 \le q \le 2$; since there is no third region 
including a line segment along the real axis, this, of course, precludes the
phenomenon of the disappearance of such a region as $a$ increases 
\cite{bcc,a}. 

The right-most point at which ${\cal B}_q$ crosses the real axis is 
given by the solution of the equation of degeneracy of leading 
terms $|\lambda_{L,1}|=|\lambda_{L,4}|$ for $0 < a < a_m$,
\beq
q_c = \frac{4(1-a)(a^2+2a+2)}{(a+1)(a+2)} \quad {\rm for} \quad 0 < a <
a_m \ ,
\label{qc}
\eeq
and is given by the solution of the equation
$|\lambda_{L,2}|=|\lambda_{L,4}|$ for $a > a_m$.
The value of $q_c$ decreases monotonically from 4 to 0 as $a$ increases from 
0 to 1. For comparison, for the $L_x \to \infty$ limits of the $L_y=2$
cyclic or M\"obius strips of the square and triangular lattices, we found
\cite{bcc,a,ta}
\beq
q_c(sq,2 \times \infty,cyc)=(1-a)(2+a) 
\label{qclad}
\eeq
for all $a$, and 
\beq
q_c(tri,2 \times \infty,cyc)=\frac{(1-a)(3+2a)}{1+a}
\label{qctri}
\eeq
for $0 \le a \le (1/4)(-3+\sqrt{17}) \simeq 0.280776$ (with a more complicated
form holding for larger $a$). 
These all have the same property of monotonically decreasing to 0 as $a$
increases from 0 to 1.

\subsubsection{Ferromagnetic range $a \ge 1$}

\begin{figure}[hbtp]
\centering
\leavevmode
\epsfxsize=2.5in
\begin{center}
\leavevmode
\epsffile{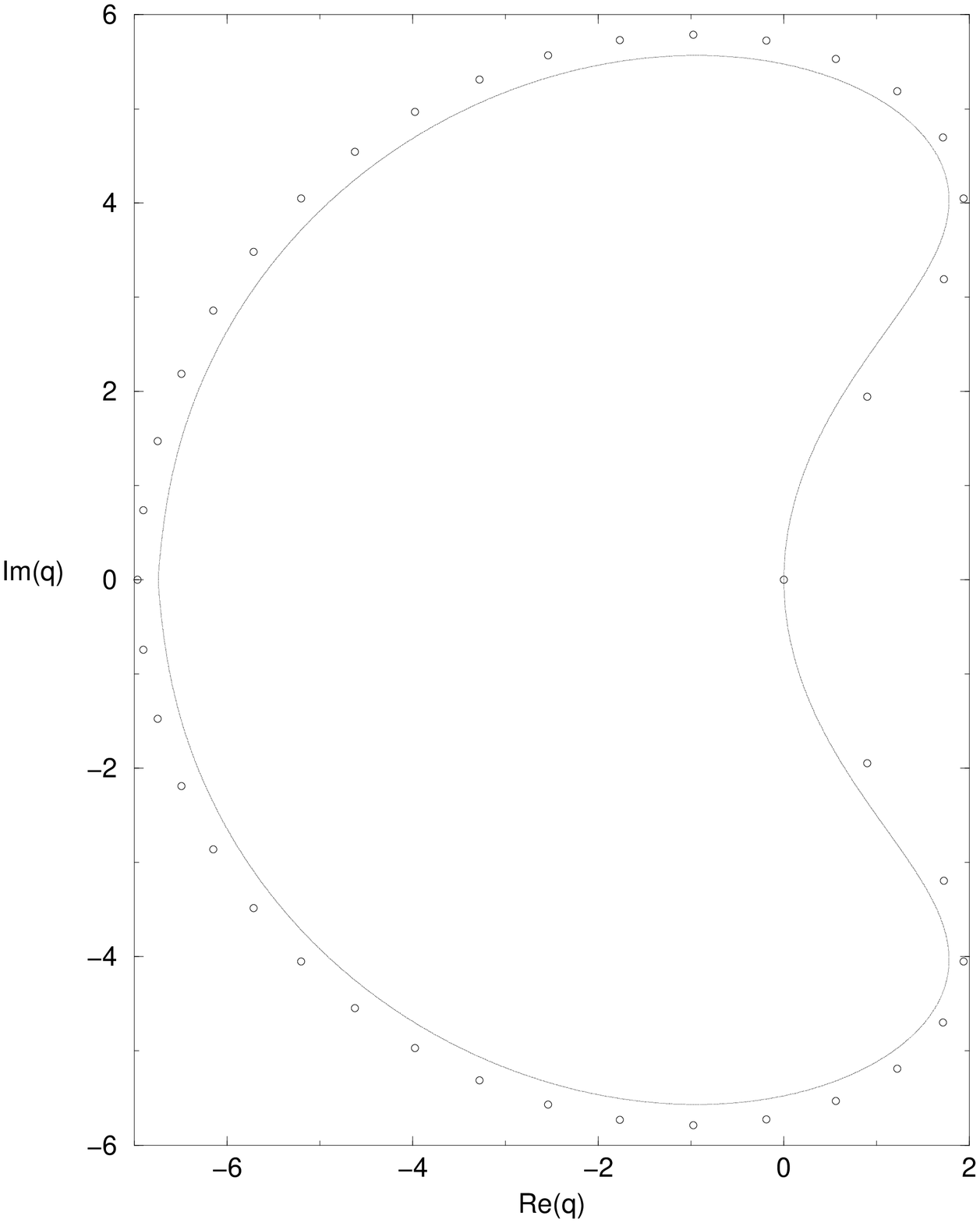}
\end{center}
\caption{\footnotesize{Locus ${\cal B}_q$ for the $n \to \infty$ limit of
the $sq_d$ strip with $a=2$.}}
\label{k4pxy2a2}
\end{figure}

For the Potts ferromagnet, as $T$ decreases from infinity, i.e. $a$ increases
above 1, the locus ${\cal B}_q$ forms a lima-bean shaped curve shown for a
typical value, $a=2$, in Fig. \ref{k4pxy2a2}.  Besides the generally present
crossing at $q=0$, the point $q_c(\{L\})$ at which ${\cal B}_q$ crosses the
real $q$ axis now occurs at negative $q$ values.  As was true of the model on
the width $L_y=2$ cyclic and M\"obius strips of the square and triangular
lattice \cite{a,ta}, for physical temperatures, the locus ${\cal B}_q$ for the
Potts ferromagnet does not cross the positive real $q$ axis.  Note that this
locus does have some support in the $Re(q) > 0$ half plane, away from the real
axis, which was also true of the analogous loci for the $L_y=2$ cyclic and
M\"obius square and triangular strips.  Finally, one could discuss the
complex-temperature range $a < 0$; however, for the sake of brevity, we shall
not do this here.

\subsection{${\cal B}_u(\{L\})$ for Fixed $q$}

We next proceed to the slices of ${\cal B}$ in the plane defined by the
temperature Boltzmann variable $u$, for given values of $q$, starting with
large $q$. In the limit $q \to \infty$, the locus ${\cal B}_u(\{L\})$ is
reduced to $\emptyset$.  This follows because for large $q$, there is only
a single dominant term, namely
\beq
\lambda_{L,4} \sim q^2 + 5qv + O(1) \quad {\rm as } \quad q \to \infty \ .
\label{lambdar1asymp}
\eeq
Note that in this case, one gets the same result whether one takes $q \to
\infty$ first and then $n=2m \to \infty$, or $n \to \infty$ and then $q
\to \infty$, so that these limits commute as regards the determination of
${\cal B}_u$.

We first consider values of $q \ne 0,1,2$, so that no noncommutativity occurs,
and $({\cal B}_u)_{nq}=({\cal B}_u)_{qn} \equiv {\cal B}_u$. As discussed
above, it is convenient to use the $u$ plane since ${\cal B}_u$ is compact in
this plane, except for the cases $q=2$, and $q=4$, whereas ${\cal B}_u$ is
noncompact because of the antiferromagnetic zero-temperature critical point at
$a=1/u=0$.  Extending the discussion in \cite{a} to the case of the strip of
the $sq_d$ lattice, we observe that the property that the singular locus ${\cal
B}_u$ passes through the $T=0$ point $u=0$ for the Potts model with $PBC_x$ but
not with $FBC_x$, i.e., with periodic, but not free, longitudinal boundary
conditions, means that the use of $PBC_x$ yields a singular locus that
manifestly incorporates the zero-temperature critical point, while this is not
manifest in ${\cal B}_u$ when calculated using $FBC_x$. 

\begin{figure}[hbtp]
\vspace{-20mm}
\centering
\leavevmode
\epsfxsize=2.5in
\begin{center}
\leavevmode
\epsffile{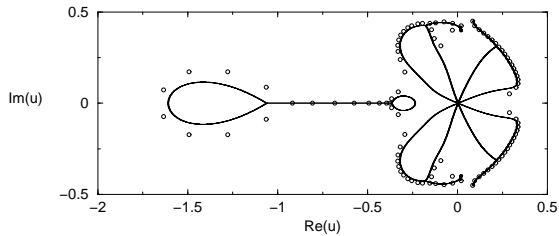}
\end{center}
\vspace{-20mm}
\caption{\footnotesize{Locus ${\cal B}_u(\{L\})$ for the $n \to \infty$
limit of the cyclic $sq_d$ strip with $q=10$. Partition function zeros are
shown for $m=20$, so that $Z$ is a polynomial of degree $e=5m=100$ in $v$
and hence, up to an overall factor, in $u$).}}
\label{k4pxy2q10}
\end{figure}

For $q=10$, the locus ${\cal B}_u$ is shown in Fig. \ref{k4pxy2q10}. Eight
curves forming four branches on ${\cal B}_u$ run into the origin, $u=0$, at the
angles given in general in eq. (\ref{thetau}). The $\lambda_{L,j}$'s that are
dominant in these phases are $\lambda_{L,2}$ and $\lambda_{L,4}$, in an
alternating manner as one makes a circuit around the origin.  The PM phase
includes the positive real $u$ axis and evidently extends all the way around
the curves forming ${\cal B}_u$. The locus ${\cal B}_u$ also includes a line
segment on the negative real $u$ axis along which $\lambda_{L,4}$ and
$\lambda_{L,5}$ are dominant and are equal in magnitude as complex conjugates
of each other.  In the two closed, complex-temperature O phases at the ends of
the line segment, the dominant term is $\lambda_{L,2}$.  As is evident in
Fig. \ref{k4pxy2q10}, although the complex-temperature (Fisher \cite{fisher})
zeros computed for a long finite strip lie reasonably close to the asymptotic
$L_x \to \infty$ curve ${\cal B}_u$ in general, these zeros have lower density
on the curves running into the origin.

\begin{figure}[hbtp]
\centering 
\leavevmode
\epsfxsize=2.5in
\begin{center}
\leavevmode
\epsffile{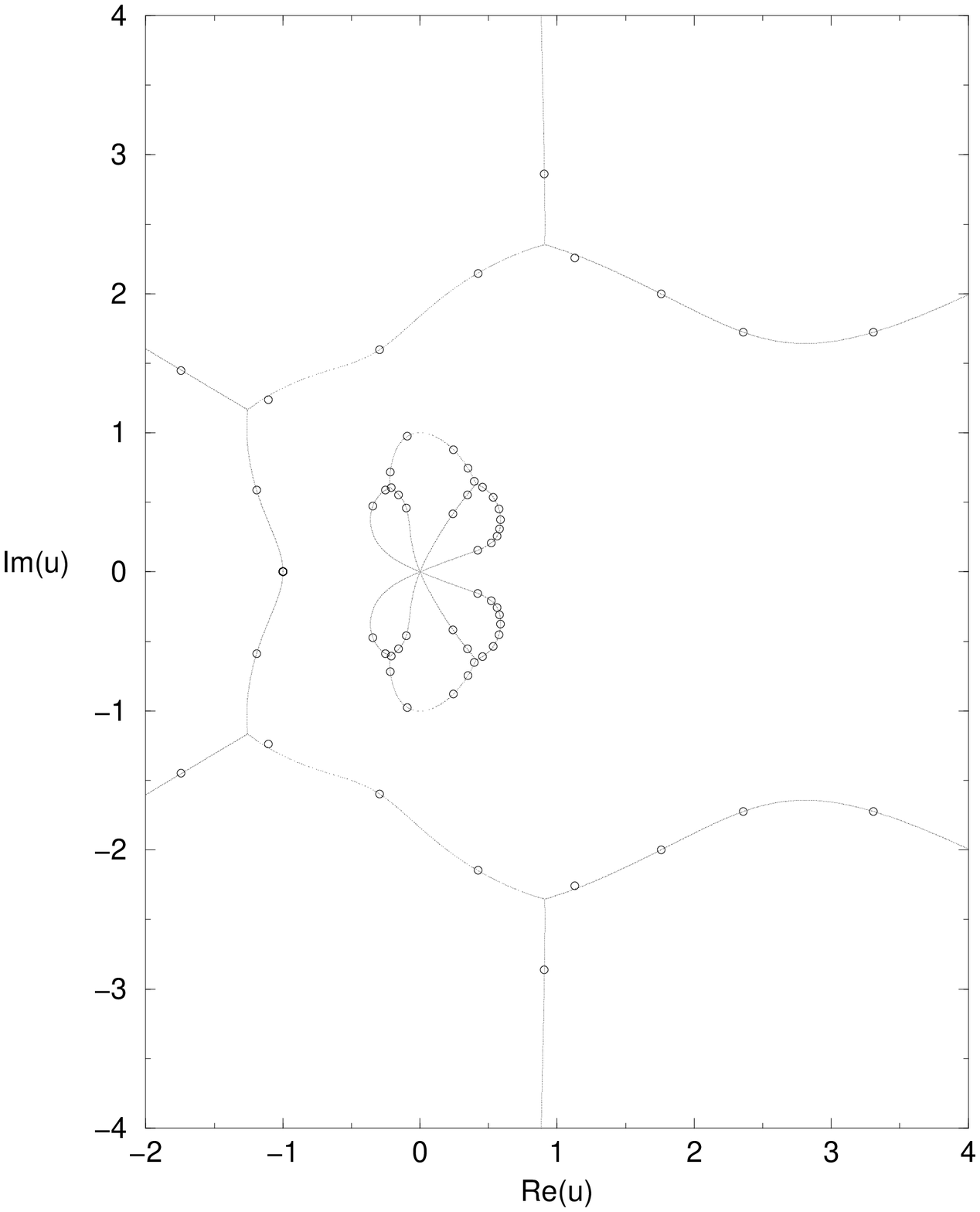}
\end{center}
\caption{\footnotesize{Locus ${\cal B}_u(\{L\})$: same as Fig. 
\ref{k4pxy2q10} for $q=2$.}}
\label{k4pxy2q2}
\end{figure}

\begin{figure}[hbtp]
\centering
\leavevmode
\epsfxsize=2.5in
\begin{center}
\leavevmode
\epsffile{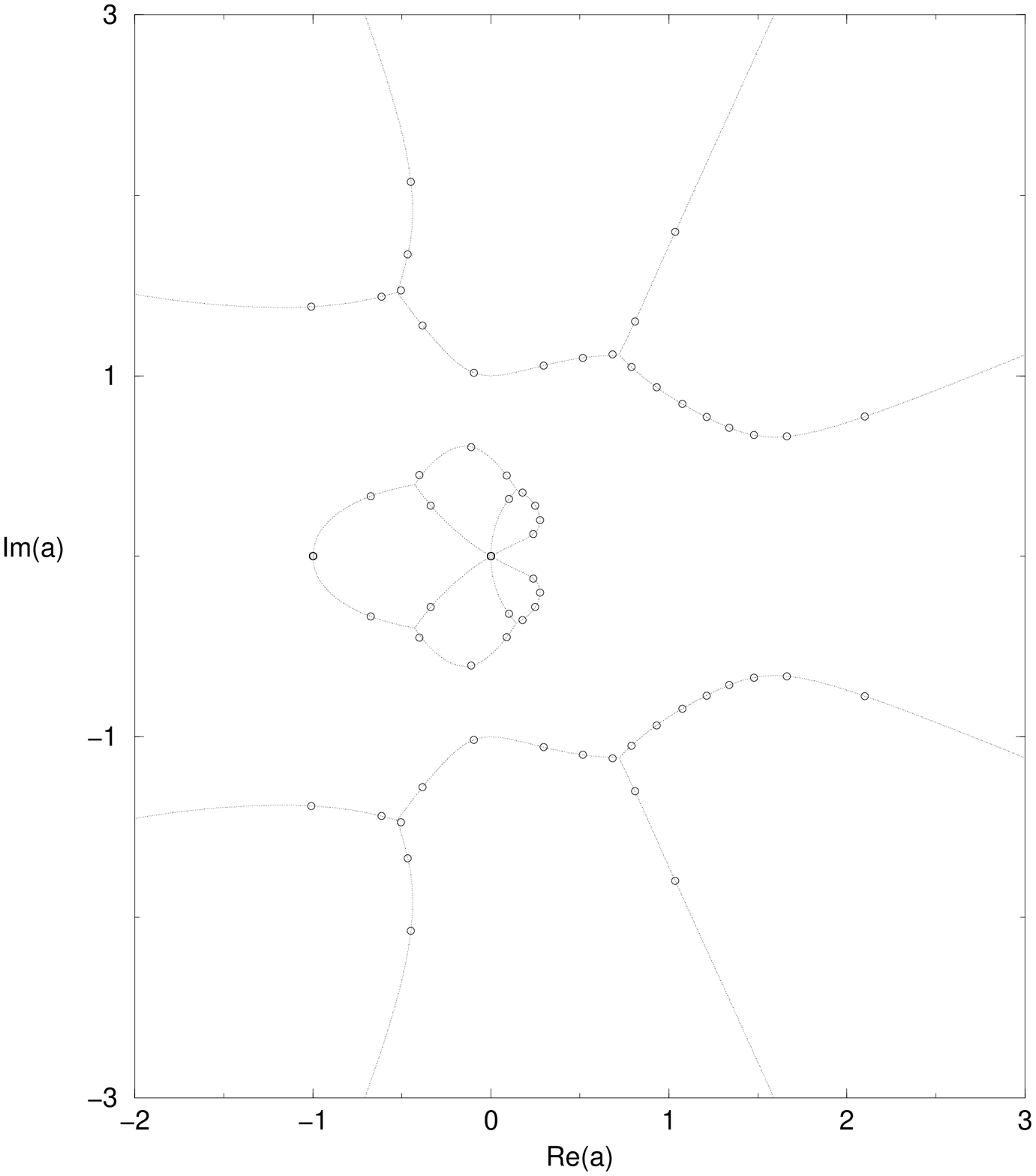}
\end{center}
\caption{\footnotesize{Locus ${\cal B}_a$ for the $n \to \infty$ limit of
the $sq_d$ strip with $q=2$.}}
\label{k4pxy2aq2}
\end{figure}

For the $q=2$ Ising case, the locus $({\cal B}_u)_{nq}$ is shown in
Fig. \ref{k4pxy2q2}.  One sees that, in addition to the eight curves
intersecting at the ferromagnetic zero-temperature critical point $u=0$, there
is evidently another $O$ phase that includes the negative real axis for $u <
-1$, and two complex conjugate pairs of $O$ phases extending toward the upper
and lower left, and upper and lower right. The complex-temperature phases in
the vicinity of $u=0$ were determined above after eq. (\ref{thetau}).  Since
the Ising antiferromagnet on the $L_x \to \infty$ limit of this strip has a 
zero-temperature critical point, it is useful to display the
singular locus ${\cal B}_a$ in the $a$ plane; here the critical point occurs
at $a=0$.  We show this in Fig. \ref{k4pxy2aq2}.  Six curves forming three
branches on ${\cal B}_a$ pass through $a=0$ at
the angles given in eqs.  (\ref{phia}).

\begin{figure}[hbtp]
\centering 
\leavevmode
\epsfxsize=2.5in
\begin{center}
\leavevmode
\epsffile{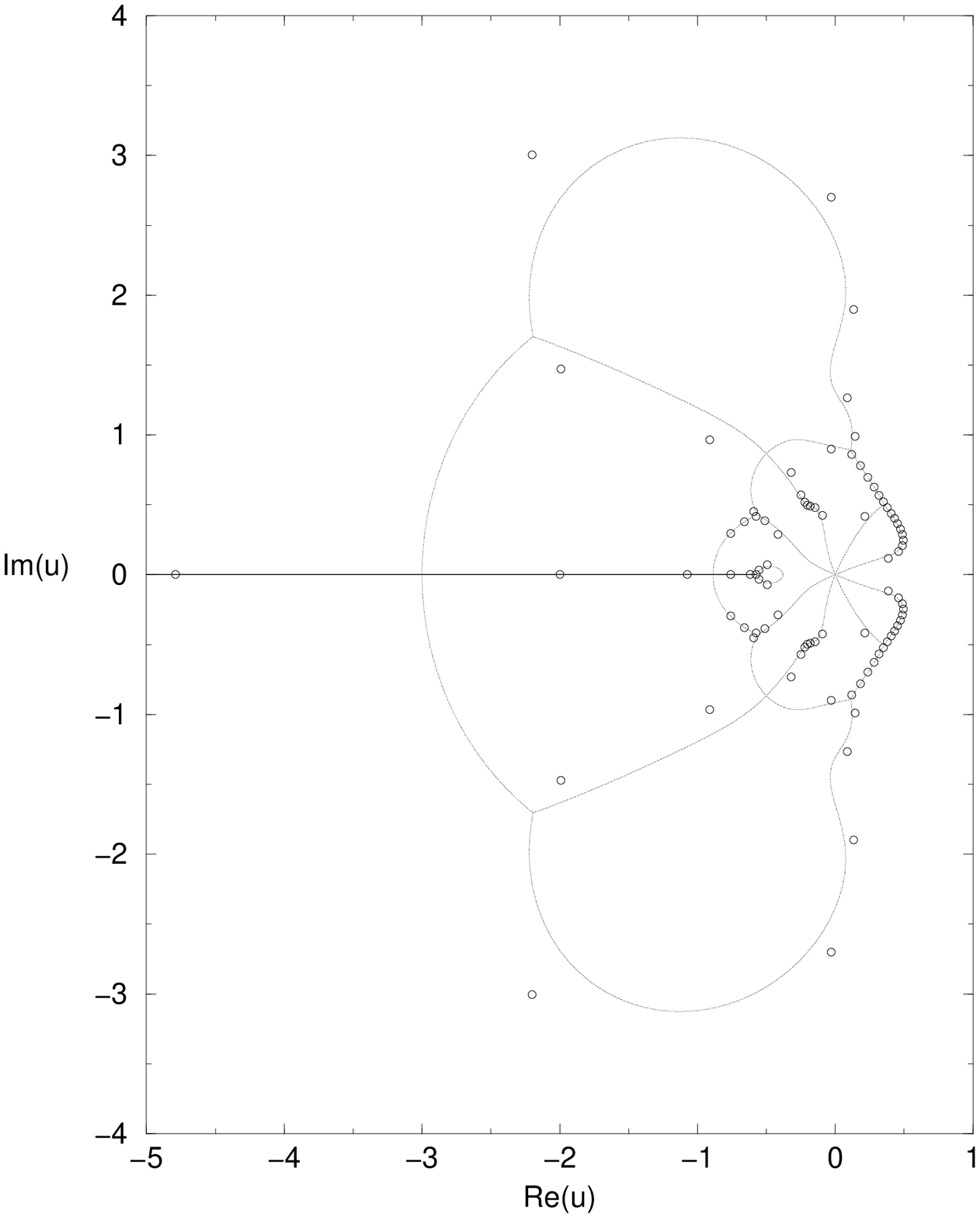}
\end{center}
\caption{\footnotesize{Locus ${\cal B}_u(\{L\})$: same as Fig. 
\ref{k4pxy2q10} for $q=3$.}}
\label{k4pxy2q3}
\end{figure}

For $q=3$, the locus ${\cal B}_u$ is shown in Fig. \ref{k4pxy2q3}. In this
case, in addition to the eight phases that are contiguous at $u=0$, there is a
line segment on the real $u$ axis from $-0.573$ to $-\infty$ and a small $O$
phase at the right end of this segment. There are also two complex conjugate
pairs of $O$ phases with the property that one pair is contiguous with the
$O_2$ and $O_2^*$ phases while another pair is contiguous with the $O_3$,
$O_3^*$ and $O_4$ phases.  Normally, the fact that part of ${\cal B}$ extends
to the point $a=0$, as is the case here, means that for the given value of $q$
the Potts antiferromagnet has a zero-temperature critical point.  However, here
one encounters noncommutativity in the definition of the free energy; if one
sets $a=0$ (i.e., $v=-1$) first, then $\lambda_{L,5}$ vanishes and there is no
such semi-infinite line segment on ${\cal B}$.

\begin{figure}[hbtp]
\centering 
\leavevmode
\epsfxsize=2.5in
\begin{center}
\leavevmode
\epsffile{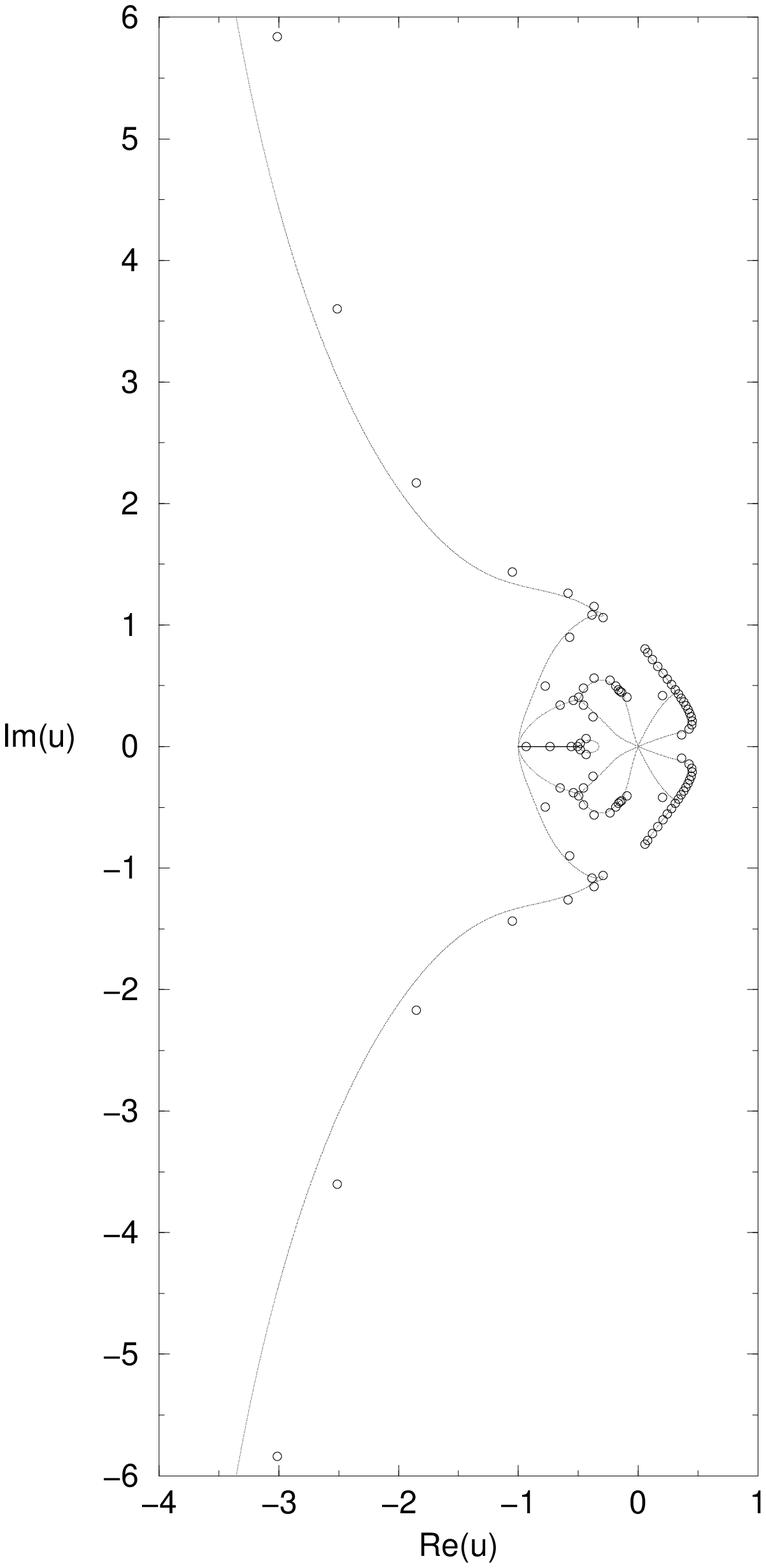}
\end{center}
\caption{\footnotesize{Locus ${\cal B}_u(\{L\})$: same as Fig. 
\ref{k4pxy2q10} for $q=4$.}}
\label{k4pxy2q4}
\end{figure}

\begin{figure}[hbtp]
\centering
\leavevmode
\epsfxsize=2.5in
\begin{center}
\leavevmode
\epsffile{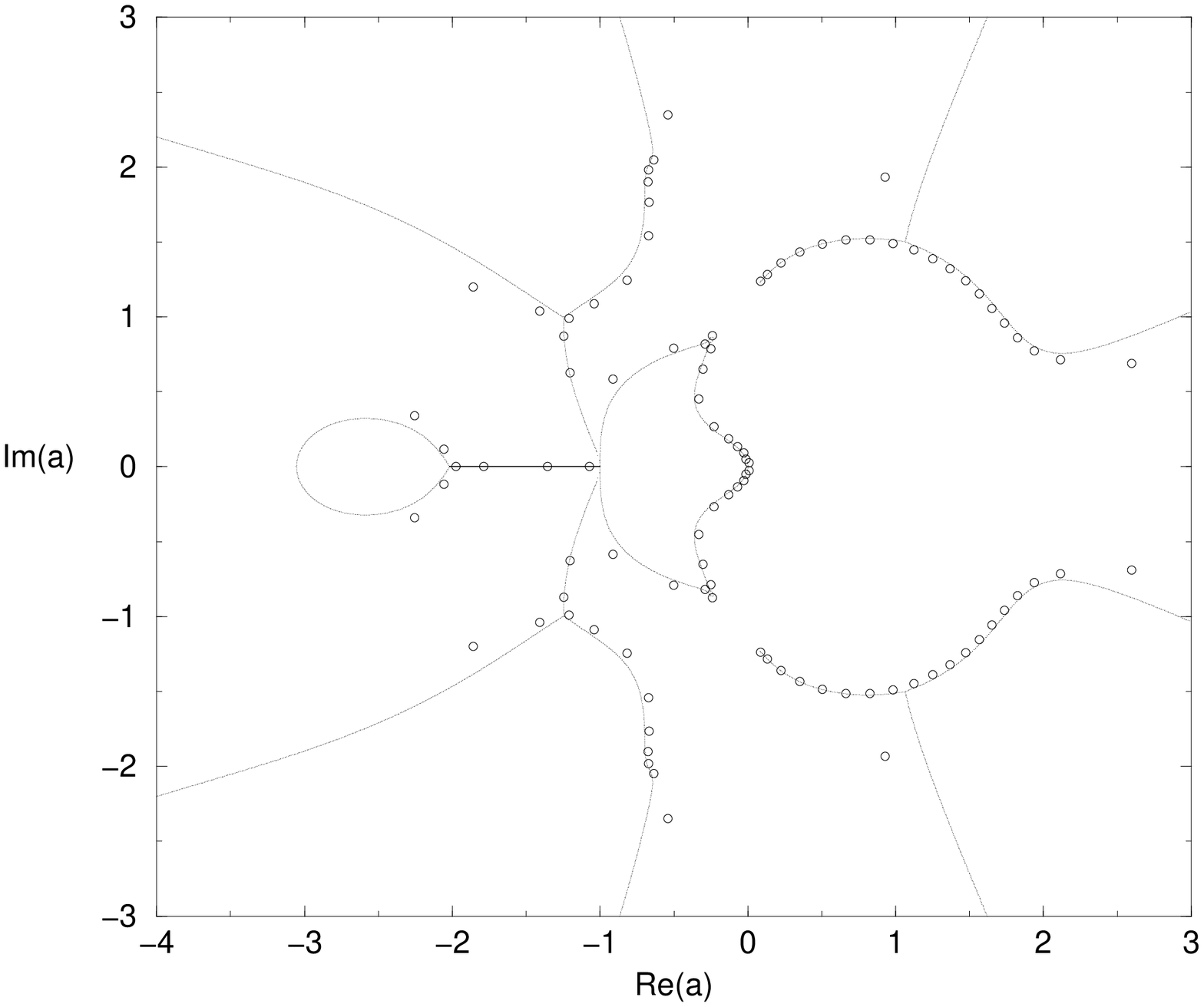}
\end{center}
\caption{\footnotesize{Locus ${\cal B}_a$ for the $n \to \infty$ limit of
the $sq_d$ strip with $q=4$.}}
\label{k4pxy2aq4}
\end{figure}

For $q=4$, the locus ${\cal B}_u$ is shown in Fig. \ref{k4pxy2q4}. There is
evidently another $O$ phase that includes the negative real axis for $u <
-1$. The line segment $-1 < u < -0.5$ on the negative real $u$ axis and the
small $O$ phase at the right end of it are also present here.  To show the
behavior in the vicinity of the zero-temperature critical point of this $q=4$
Potts antiferromagnet, we display the singular locus ${\cal B}_a$ in the $a$
plane in Fig. \ref{k4pxy2aq4}.  The locus ${\cal B}_a$ passes vertically
through the origin, $a=0$ and then curves into the $Re(a) < 0$ half-plane. 

\subsection{Thermodynamics of the Potts Model on the $L_y=2$ Strip of the
$sq_d$ Lattice}

\subsubsection{Ferromagnetic Case}

\begin{figure}[hbtp]
\centering
\leavevmode
\epsfxsize=4.0in
\begin{center}
\leavevmode
\epsffile{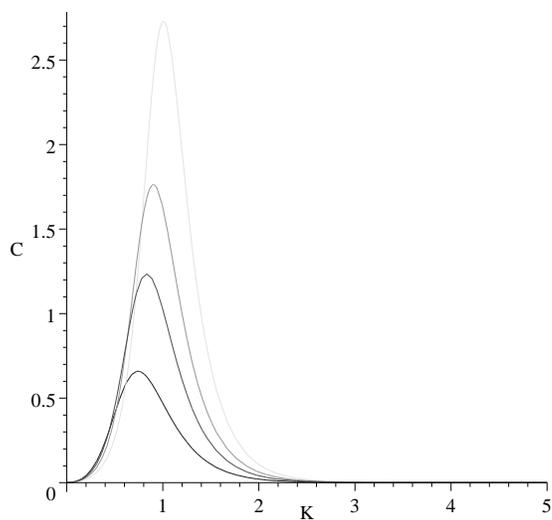}
\end{center}
\vspace{-40mm}
\caption{\footnotesize{Specific heat (with $k_B \equiv 1$) for the Potts
ferromagnet on the infinite-length, width $L_y=2$ strip of the $sq_d$ 
lattice, as a function of $K=J/(k_BT)$.  Going from bottom to top in order
of the heights of the maxima, the curves are for $q=2,3,4,6$.}}
\label{cfmsqdstrip}
\end{figure}

The Potts ferromagnet (with real $q > 0$) on an arbitrary
graph has $v > 0$ so, as is clear from eq. (\ref{cluster}), the partition
function satisfies the constraint of positivity.  In contrast, the
specific heat $C$ is positive for the model on the (infinite-length limit
of the) $L_y=2$ $sq_d$ strip if and only if $q > 1$. For
$q=1$, $f_{nq}=2.5\ln a = 2.5K$ and $C$ vanishes identically.  Since a
negative specific heat is unphysical, we therefore restrict to real $q \ge
1$. For general $q$ in this range, the reduced free energy is given for
all temperatures by $f=(1/2)\ln \lambda_{S,1}$ as in (\ref{fstrip}). 
Recall that $\lambda_{S,1} \equiv \lambda_{L,4}$.  It is straightforward
to obtain the internal energy $U$ and specific heat from this free energy;
since the expressions are somewhat complicated, we do not list them here.
We show a plot of the specific heat (with $k_B=1$) in Fig. 
\ref{cfmsqdstrip}.  One can observe that the value of the maximum is a
monotonically increasing function of $q$.

The high-temperature expansion of $U$ is
\beq
U=-\frac{5J}{2q}\biggl [ 1+\frac{(q-1)}{q}K + O(K^2) \biggr ] \ .
\label{ustriphigh}
\eeq
For the specific heat we have
\beq
C=\frac{5k_B(q-1)K^2}{2q^2}\biggl [ 1 + \frac{(5q+14)}{5q}K + O(K^2)
\biggr ]
\ .
\label{cstriphigh}
\eeq
The low-temperature expansions ($K \to \infty$) are
\beq
U = J\Biggl [ -\frac{5}{2} + 2(q-1)e^{-4K}\biggl [ 1 + 5e^{-K} +
7(q-2)e^{-2K} + O(e^{-3K}) \biggr ] \Biggr ] \quad {\rm as} \quad K \to
\infty
\label{ustriplowfm}
\eeq
and
\beq
C = 8k_BK^2(q-1)e^{-4K}\biggl [1+\frac{25}{4}e^{-K} +
\frac{49}{4}(q-2)e^{-3K} + O(e^{-4K})\biggr ] \quad {\rm as} \quad K \to
\infty
\label{cstriplowfm}
\eeq

Comparing with our corresponding calculations for the ($L_x \to \infty$ limits
of the) strips of the square and triangular lattices with the same $L_y=2$
width, we can remark on some common features.   In all of these cases, the
high-temperature expansions have the leading forms
\beq
U=-\frac{\Delta J}{2q}\biggl [ 1+\frac{(q-1)}{q}K + O(K^2) \biggr ]
\label{ustriphighgen}
\eeq
where we recall that the coordination number is 
$\Delta=3,4$ and 5 for these infinite-length strips of the
square, triangular, and $sq_d$ lattices with width 2. (In the infinite-length
limit, the longitudinal boundary conditions do not affect the coordination
number.) Further, 
\beq
C=\frac{\Delta k_B(q-1)K^2}{2q^2}\biggl [ 1 + O(K) \biggr ]
\ .
\label{cstriphighgen}
\eeq
For the low-temperature expansions for these strips, 
\beq
U = J\Biggl [ -\frac{\Delta}{2} + O((q-1)e^{-(\Delta-1)K}) \Biggr ] 
\quad {\rm as} \quad K \to \infty
\label{ustriplowfmgen}
\eeq
and
\beq
C \propto k_BK^2(q-1)e^{-(\Delta-1)K}\biggl [ 1 + O(e^{-K}) \biggr ]
\quad {\rm as} \quad K \to \infty \ . 
\label{cstriplowfmgen}
\eeq

In general, the ratio $\rho$ of the largest subdominant to the dominant
$\lambda_j$'s determines the asymptotic decay of the connected spin-spin
correlation function and hence the correlation length
\beq
\xi = -\frac{1}{\ln \rho}
\label{xi}
\eeq
Since $\lambda_{L,4}$ and $\lambda_{L,2}$ are the dominant and leading
subdominant $\lambda_j$'s, respectively, we have
\beq
\rho_{FM}=\frac{\lambda_{L,2}}{\lambda_{L,4}}
\label{rho}
\eeq
and hence for the ferromagnetic zero-temperature critical point we find
that   
the correlation length diverges, as $T \to 0$, as
\beq
\xi_{FM} \sim q^{-1}e^{4K} \quad {\rm as} \quad K \to \infty
\label{xit}
\eeq
Comparing with the divergences in the correlation length at the ferromagnetic 
$T=0$ critical point that we have
calculated for the infinite-length limits of the square and triangular strips
with the same $L_y=2$ width \cite{a,ta}, we see that all of these can be fit 
by the formula
\beq
\xi_{FM} \sim q^{-1}e^{(\Delta-1)K} \quad {\rm as} \quad K \to \infty \ .
\label{xifmgen}
\eeq

\subsubsection{Antiferromagnetic Case}

In this section we first restrict to the real range $q \ge 4$ and the
additional integer values $q=2$ (Ising case) and $q=3$ where the Potts
antiferromagnet exhibits physically acceptable behavior and then consider the
remaining interval $0 < q < 4$, $q \ne 2,3$, where it exhibits unphysical
properties.  For $q \ge 4$, the free energy is given for all temperatures by
(\ref{fstrip}), as in the ferromagnetic case but with $J$ negative, and is the
same independent of the different longitudinal boundary conditions, as is
necessary for there to exist a thermodynamic limit.

\begin{figure}[hbtp]
\centering
\leavevmode
\epsfxsize=4.0in
\begin{center}
\leavevmode
\epsffile{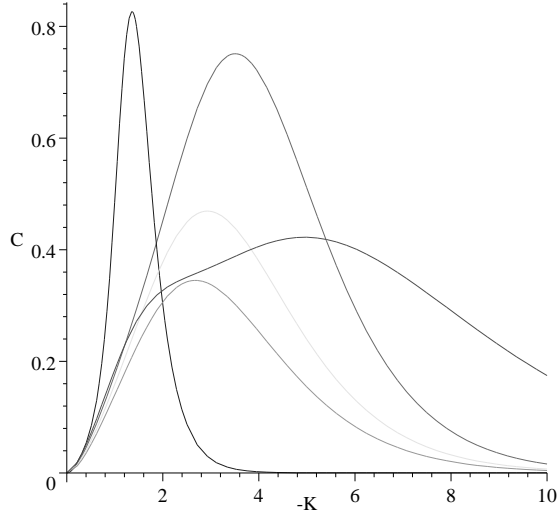}
\end{center}
\vspace{-40mm}
\caption{\footnotesize{Specific heat (with $k_B \equiv 1$) for the Potts  
antiferromagnet on the infinite-length, width $L_y=2$ strip of the $sq_d$
lattice, as a function of $-K = -J/(k_BT)$.  Going downward in order of
the heights of the maxima, the curves are for $q=2,4,5,3,6$.}}
\label{cafmsqdstrip}
\end{figure}

We show plots of the specific heat, for several values of $q$, for the
Potts antiferromagnet on the (infinite-length limit of the) $L_y=2$ strip
of the $sq_d$ lattice in Fig \ref{cafmsqdstrip}.  In contrast to the
ferromagnetic case, where the positions of the maxima of $C$ in the variable
$K$ increase monotonically as functions of $q$, the positions of the maxima of
$C$ in the antiferromagnetic  do not have a monotonic dependence on $q$; 
they occur at $K=-0.82$ for $q=2$, $K=-0.42$ for $q=3$ and $K=-0.75$ for $q=4$,
after which the values of the maxima occur at smaller values of $-K$ with 
increasing integral $q$, e.g., $K=-0.47$ for $q=5$ and $K=-0.34$ for $q=6$).  
Recall that the antiferromagnetic Potts model with $q=2$ and $q=3$ involves
frustration, and one can observe a rather different behavior in the specific
heat for these values of $q$ as contrasted with $q \ge 4$ in the plot. 

The high-temperature expansions of $U$ and $C$ are given by 
(\ref{ustriphigh}) and (\ref{cstriphigh}); more generally, these
expansions also apply in the range $0 < q < 4$.  As discussed above, the
Ising case $q=2$ is one of the cases where one must take account of
noncommutativity in the definition of the free energy and hence of
thermodynamic quantities.  If one sets $q=2$ first and then takes $n \to
\infty$, then $f=f_{nq}=(1/2)\ln \lambda_{L,4}(q=2)$ where
$\lambda_{L,4}(q=2)$ was given in eq. (\ref{lam45q2}), and the
low-temperature expansions are
\beq
U(q=2) = -\frac{J}{2}\biggl [ 1 + 12e^{3K} + 36e^{4K} + 
80e^{5K} + O(e^{6K}) \biggr ] \quad {\rm as} \quad K \to -\infty
\label{ustriplowafmising}
\eeq
and
\beq
C(q=2) = 18k_BK^2e^{3K}\biggl [ 1 + 4e^K + \frac{100}{9}e^{2K} +
\frac{72}{9}e^{3K} + O(e^{4K}) \biggr ]
\quad {\rm as} \quad K \to -\infty \ .
\label{cstriplowfmising}
\eeq

For $q=3$, if one sets $q=3$ first and then takes $n \to
\infty$, then $f=f_{nq}=(1/2)\ln \lambda_{L,4}(q=3)$, and the
low-temperature expansions are
\beq
U(q=3) = -\frac{J}{4}\biggl [ 1 +\frac{3\sqrt{2}}{2}e^{K/2} - 4e^{K} + 
\frac{57\sqrt{2}}{8}e^{3K/2} + O(e^{2K}) \biggr ] \quad {\rm as} \quad K
\to -\infty
\label{ustriplowafmq3}
\eeq
and
\beq
C(q=3) = \frac{3\sqrt{2}}{16}k_BK^2e^{K/2}\biggl [ 1 -
\frac{8\sqrt{2}}{3}e^{K/2} + \frac{57}{4}e^{K} -
16\sqrt{2}e^{3K/2} + O(e^{2K}) \biggr ]
\quad {\rm as} \quad K \to -\infty \ .
\label{cstriplowfmq3}
\eeq

For the range $q \ge 4$, the low-temperature expansions are given by
\beq U =
\frac{(-J)e^K}{2(q-3)^2}\biggl [ (5q-13) -
\frac{29q^3-190q^2+405q-276}{(q-3)^2(q-2)} e^K +
O(e^{2K}) \biggr ] \quad {\rm as} \quad K \to -\infty
\label{ustriplowafm}
\eeq
and
\beq
C = \frac{k_BK^2e^K}{2(q-3)^2}\biggl [ (5q-13) -
\frac{2(29q^3-190q^2+405q-276)}{(q-3)^2(q-2)} e^K + O(e^{2K}) \biggr ]
\quad {\rm as} \quad K \to -\infty \ .
\label{cstriplowafm}
\eeq
Note that for the antiferromagnetic case, $U(T=0)=0$ for $q \ge 4$, but
$U(T=0)=-J/2=+|J|/2$ for $q=2$ and $U(T=0)=-J/4=+|J|/4$ for $q=3$. The
vanishing value of $U$ at $T=0$ for
$q \ge 4$ means that the Potts model can achieve its
preferred ground state for this range of $q$, while the nonzero value of
$U(T=0)$ for the Ising and $q=3$ antiferromagnet is a consequence of the
frustration
that is present in this case.  
Note that the apparent divergences that occur as $q \to
2$ and $q \to 3$ in eqs. (\ref{ustriplowafm}) and (\ref{cstriplowafm}) are
not actually reached here since these expressions apply only in the region
$q \ge 4$ (the discrete integral cases $q=2$ and $q=3$ were dealt with
above).

For the zero-temperature critical points in the $q=2$ and $q=4$ Potts
antiferromagnet,
\beq
\rho_{AFM,q=2,4}=\frac{\lambda_{L,2}}{\lambda_{L,4}}
\label{rhoafmq2}
\eeq
Using the respective expansions (\ref{lam4q2taylor})-(\ref{lam5q2taylor})
and (\ref{lam2q4taylor})-(\ref{lam5q4taylor}), we find that the
correlation lengths defined as in (\ref{xi}) diverges, as $T \to 0$, as
\beq
\xi_{AFM,q=2} \sim e^{-3K} \ , \quad {\rm as} \quad K \to -\infty
\label{xiafmq2}
\eeq
and
\beq
\xi_{AFM,q=4} \sim e^{-K} \ , \quad {\rm as} \quad K \to -\infty \ . 
\label{xiafmq4}
\eeq

Next, we consider the range of real $0 < q < 4$ aside from the integral
case $q=2,3$.  The first pathology is that the Potts antiferromagnet 
on the infinite-length limit of the cyclic $L_y=2$ $sq_d$ strip 
has a phase transition at a finite temperature, call it $T_{p,L}$, 
while, in contrast, if one uses
free boundary conditions, then either (i) there is no phase transition at any 
finite temperature, for $3 < q < 4$ or $1 < q < 2$, or (ii) there is a phase
transition at a finite temperature $T_{p,S}$ for $2 < q < 3$ or $0 < q < 1$, 
but $T_{p,L} \ne T_{p,S}$, so that
there is no well-defined thermodynamic limit for the Potts antiferromagnet
with non-integral $q$ in the interval $0 < q < 4$.  
The Ising case $q=2$ and $q=3$ case have been dealt
with in the preceding subsection.  Concerning the value $q=1$, as
discussed earlier, one encounters noncommutativity in defining the free
energy.  If one takes $q=1$ to start with and then $n \to \infty$, the
thermodynamic limit does exist, independent of boundary conditions, and
$f=f_{nq}=2.5K$, $U=-2.5J=2.5|J|$, and $C=0$.  If one starts with $q \ne
1$, takes $n \to \infty$, calculates $f_{qn}$, and then takes $q \to 1$,
the thermodynamic limit does not exist since the result differs depending
on whether one uses free longitudinal boundary conditions or cyclic 
longitudinal boundary conditions. In the high-temperature phase, 
$f_{qn}=(1/2)\ln \lambda_{L,4}$, independent of longitudinal boundary
conditions, but in the low-temperature phase, the expression for $f_{qn}$
is different for the open and cyclic strips.  There are also other
unphysical properties, such as a negative specific heat and a negative
partition function for certain ranges of temperature.

Acknowledgment: The research of R. S. was supported in part at Stony Brook by
the U. S. NSF grant PHY-97-22101 and at Brookhaven by the U.S. DOE contract
DE-AC02-98CH10886.\footnote{\footnotesize{Accordingly, the U.S. government
retains a non-exclusive royalty-free license to publish or reproduce the
published form of this contribution or to allow others to do so for
U.S. government purposes.}}

\section{Appendix}

\subsection{General} 

The Potts model partition function $Z(G,q,v)$ is related to the Tutte
polynomial $T(G,x,y)$ as follows.
The graph $G$ has vertex set $V$ and edge set $E$,
denoted $G=(V,E)$.  A spanning subgraph $G^\prime$ is defined as a subgraph
that has the same vertex set and a subset of the edge set:
$G^\prime=(V,E^\prime)$ with $E^\prime \subseteq E$.  The Tutte polynomial
of $G$, $T(G,x,y)$, is then given by \cite{tutte1}-\cite{tutte3}
\beq
T(G,x,y)=\sum_{G^\prime \subseteq G} (x-1)^{k(G^\prime)-k(G)}
(y-1)^{c(G^\prime)}
\label{tuttepol}
\eeq
where $k(G^\prime)$, $e(G^\prime)$, and $n(G^\prime)=n(G)$ denote the number
of components, edges, and vertices of $G^\prime$, and
\beq
c(G^\prime) = e(G^\prime)+k(G^\prime)-n(G^\prime)
\label{ceq}
\eeq
is the number of independent circuits in $G^\prime$ (sometimes called the
co-rank of $G^\prime$).   Note that the first factor can also be written as
$(x-1)^{r(G)-r(G^\prime)}$, where
\beq
r(G) = n(G)-k(G)
\label{rank}
\eeq
is called the rank of $G$.  The graphs $G$ that we consider here are
connected, so that $k(G)=1$.  Now let
\beq
x=1+\frac{q}{v}
\label{xdef}
\eeq
and
\beq
y=a=v+1
\label{ydef}
\eeq
so that
\beq
q=(x-1)(y-1) \ . 
\label{qxy}
\eeq
Then
\bigskip
\beq
Z(G,q,v)=(x-1)^{k(G)}(y-1)^{n(G)}T(G,x,y) \ .
\label{ztutte}
\eeq
Note that the chromatic polynomial is a special case of the Tutte polynomial:
\beq
P(G,q)=q^{k(G)}(-1)^{k(G)+n(G)}T(G,x=1-q,y=0)
\label{tprel}
\eeq
(recall eq. (\ref{zp})).

Corresponding to the form (\ref{zgsum}) we find that the Tutte polynomial for
recursively defined graphs comprised of $m$ repetitions of some subgraph has
the form
\beq
T(G_m,x,y) = \sum_{j=1}^{N_\lambda} c_{T,G,j}(\lambda_{T,G,j})^m
\label{tgsum}
\eeq

\subsection{$sq_d$ Strip with Free Longitudinal Boundary Conditions}

The generating function representation for the Tutte polynomial for the open
strip of the $sq_d$ lattice comprised of $m+1$ squares with edges joining
the lower-left to upper-right vertices and the upper-left to lower-right
vertices of each square, denoted $S_m$, is
\beq
\Gamma_T(S_m,x,y;z) = \sum_{m=0}^\infty T(S_m,x,y)z^m \ .
\label{gammatfbc}
\eeq
We have
\beq
\Gamma_T(S,x,y;z) = \frac{{\cal N}_T(S,x,y;z)}{{\cal D}_T(S,x,y;z)}
\label{gammas}
\eeq
where
\beq
{\cal N}_T(S,x,y;z)=A_{T,S,0}+A_{T,S,1}z
\label{numts}
\eeq
with
\beq
A_{T,S,0}=2x+4xy+3x^2+2y+3y^2+y^3+x^3
\label{as0tut}
\eeq
\beq
A_{T,S,1}=xy(y+y^2+x-2xy^2+xy+x^2-2yx^2-x^2y^2)
\label{as1tut}
\eeq
and
\beq
{\cal D}_T(S,x,y;z) = \prod_{j=1}^2 (1-\lambda_{T,S,j}z)
\label{dents}
\eeq
with
\beq
\lambda_{T,S,(1,2)} = \frac{1}{2}\biggl [ x(x+3)+y(y^2+2y+3)+2 \pm 
\sqrt{R_T} \ \biggr ]
\label{lamstut12}
\eeq
where
\beqs
R_T & = & 4+12x+12y+22xy+13x^2+21y^2+6x^3+20y^3+16xy^2 \cr\cr
& & +10x^2y+x^4+10y^4-4x^2y^2-2y^3x+4y^5-2x^2y^3+y^6 \ . 
\label{rt12}
\eeqs

The corresponding closed-form expression is given by the general formula from
\cite{hs}, as applied to Tutte, rather than chromatic, polynomials, namely
\beq
T(S_m,x,y)=\biggl [
\frac{A_{T,S,0}\lambda_{T,S,1}+A_{T,S,1}}{\lambda_{T,S,1}-\lambda_{T,S,2}}
\biggr ] (\lambda_{T,S,1})^m + \biggl [
\frac{A_{T,S,0}\lambda_{T,S,2}+A_{T,S,1}}
{\lambda_{T,S,2}-\lambda_{T,S,1}} \biggr ] (\lambda_{T,S,2})^m \ .
\label{tssumform}
\eeq
It is easily checked that this is a symmetric
function of the $\lambda_{S,j}$, $j=1,2$.

\subsection{Cyclic and M\"obius Strips}

We write the Tutte polynomials for the cyclic and M\"obius strips of the
$sq_d$ lattice with width $L_y=2$ as 
\beq
T(L_m,x,y) = \sum_{j=1}^5 c_{T,L,j}(\lambda_{T,L,j})^m
\label{tlxy}
\eeq
where it is convenient to extract a common factor from the coefficients:
\beq
c_{T,L,j} \equiv \frac{\bar c_{T,L,j}}{x-1} \ . 
\label{cbar}
\eeq
Of course, although the individual terms contributing
to the Tutte polynomial are thus rational functions of $x$ rather than
polynomials in $x$, the full Tutte polynomial is a polynomial
in both $x$ and $y$.  We have
\beq
\lambda_{T,L,1}=2
\label{lam1tut}
\eeq
and, with
\beq
T_{T,23}=y^3+2y^2+3y+2x+4
\label{t23tut}
\eeq
\beq
R_{T,23}=16+16x+32y+12xy+33y^2-4xy^3+10y^4+4x^2+20y^3+4y^5+y^6
\label{r23tut}
\eeq
the results 
\beq
\lambda_{T,L,(2,3)} = \frac{1}{2}\biggl [ T_{T,23} \pm \sqrt{R_{T,23}} \ \biggr
] 
\label{lam23tut}
\eeq
and 
\beq
\lambda_{T,L,(4,5)} = \lambda_{T,S,(1,2)}
\label{lam45tut}
\eeq

The coefficients are 
\beq
\bar c_{T,L,1} = \frac{1}{2}(x-1)(y-1)(xy-x-y-2)
\label{c1tutt}
\eeq
\beq
\bar c_{T,L,2} = \bar c_{T,L,3} = xy-x-y
\label{c234tutt}
\eeq
\beq
\bar c_{T,L,4} = \bar c_{T,L,5} = 1 \ . 
\label{c45tutt}
\eeq
These are symmetric under interchange of $x \leftrightarrow y$, which is a
consequence of the fact that the $c_{L,j}$ are functions only of $q$. 

\subsection{Special Values of Tutte Polynomials for Strips of the $sq_d$
Lattice}

For a given graph $G=(V,E)$, at certain special values of the arguments $x$ and
$y$, the Tutte polynomial $T(G,x,y)$ yields quantities of basic graph-theoretic
interest \cite{tutte3}-\cite{boll}, \cite{wu77}.  We
recall some definitions: a spanning subgraph $G^\prime=(V,E^\prime)$ of $G$
is a graph with the same vertex set $V$ and a subset of the edge set,
$E^\prime \subseteq E$.  Furthermore, a tree is a graph with no cycles, and a
forest is a graph containing one or more trees.  Then the number of spanning
trees of $G$, $N_{ST}(G)$, is
\beq
N_{ST}(G)=T(G,1,1) \ ,
\label{t11}
\eeq
the number of spanning forests of $G$, $N_{SF}(G)$, is
\beq
N_{SF}(G)=T(G,2,1) \ ,
\label{t21}
\eeq
the number of connected spanning subgraphs of $G$, $N_{CSSG}(G)$, is
\beq
N_{CSSG}(G)=T(G,1,2) \ ,
\label{T12}
\eeq
and the number of spanning subgraphs of $G$, $N_{SSG}(G)$, is
\beq
N_{SSG}(G)=T(G,2,2) \ .
\label{t22}
\eeq

 From our calculations of Tutte polynomials, we find that
\beq
N_{ST}(S_m) = 2^{2(m+2)} \cdot 3^m 
\label{t11sm}
\eeq

\beq
N_{SF}(S_m) = (19+11\sqrt{3} \ )(9+5\sqrt{3} \ )^m
+(19-11\sqrt{3} \ )(9-5\sqrt{3} \ )^m
\label{t21sm}
\eeq

\beq
N_{CSSG}(S_m) = 
\Bigl (19 + \frac{65\sqrt{46}}{23} \ \Bigr )[2(7+\sqrt{46} \ )]^m+
\Bigl (19 - \frac{65\sqrt{46}}{23} \ \Bigr )[2(7-\sqrt{46} \ )]^m
\label{t12sm}
\eeq

\beq
N_{SSG}(S_m) = 2^{5m+6} \ .
\label{t22sm}
\eeq

For the cyclic $L_y=2$ strip of the $sq_d$ lattice, $L_m$, we first note 
that for $m \ge 3$, $L_m$ is a (proper) graph, but for $m=1$ and $m=2$, $L_m$
is not a proper graph, but instead, is a multigraph, with multiple edges (and,
for $m=1$, loops). The following formulas apply 
for all $m \ge 1$: 
\beq
N_{ST}(L_m) = 2^{2(m-1)} \cdot 3^m m
\label{t11lm}
\eeq

\beq
N_{SF}(L_m) =
\biggl [ \Bigl ( 9+5\sqrt{3} \ \Bigr )^m + \Bigl (9-5\sqrt{3} \ 
\Bigr )^m \biggl ] -
\biggl [ \Bigl (7+3\sqrt{5} \ \Bigr )^m + \Bigl ( 7-3\sqrt{5} \ 
\Bigr )^m \biggl ] 
\label{nsflm}
\eeq

\beqs
& & N_{CSSG}(L_m) = -3 \cdot 2^{m-1} + 
\biggl [ 2 \Bigl ( 7+\sqrt{46} \ \Bigr ) \biggr ]^m + 
\biggl [ 2 \Bigl ( 7-\sqrt{46} \ \Bigr ) \biggr ]^m + \cr\cr
& & \frac{m}{2}\Biggl [ \Bigl ( 3 + \frac{7\sqrt{46}}{23} \ \Bigr )
\biggl [ 2 \Bigl ( 7+\sqrt{46} \ \Bigr ) \biggr ]^{m-1} + 
\Bigl ( 3 - \frac{7\sqrt{46}}{23} \ \Bigr )
\biggl [ 2 \Bigl (7-\sqrt{46} \ \Bigr ) \biggr ]^{m-1} \Biggr ]
\label{ncssglm}
\eeqs

\beq 
N_{SSG}(L_m) = 2^{5m} \ . 
\label{nssglm}
\eeq
This result, eq. (\ref{nssglm}), is a special case of a more general
elementary theorem, namely 
\beq
N_{SSG}(G) = 2^{e(G)}
\label{nssglmg}
\eeq
This is proved by noting that a spanning subgraph $G^\prime \subseteq G$ 
consists of the same vertex set $V$ as $G$ and a subset of the edge set $E$ of
$G$. One can enumerate all such spanning subgraphs as follows: for each edge in
$E$, one has the option of including or excluding it, keeping the other edges
fixed.  This is a twofold choice for each edge, and the result (\ref{nssglmg})
therefore holds.  This general result subsumes the previous specific relations 
$N_{SSG}=2^{3m}$ and $N_{SSG}=2^{4m}$ for the cyclic strips of length $L_x=m$
and width $L_y=2$ of the square and triangular lattices. 
We recall that the number of edges is given by $e(G)=$ (i) $3m$, (ii) $4m$, 
and (iii) $5m$ for the $L_x=m$, $L_y=2$ strips of the (i) square (ii) 
triangular, (iii) $sq_d$ strips, respectively, while the number of
vertices is given by $n=v(G)=2m$ for all of these strips.  Let us denote 
$r_e=e(G)/n(G)$, whence $r_e(G)=$ (i) 3/2, (ii) 2, (iii) $5/2$ and 
$N_{SSG}(L_{G,m}) = 2^{r_e(G)n}$ for these strips. 

Since $T(G_m,x,y)$ grows exponentially as $m
\to \infty$ for the families $G_m=S_m$ and $L_m$ for $(x,y)=(1,1)$, 
(2,1), (1,2), and (2,2), one defines the corresponding constants
\beq
z_{set}(\{G\}) = \lim_{n(G) \to \infty} n(G)^{-1} \ln N_{set}(G) \ , \quad 
set = ST, \ SF, \ CSSG, \ SSG
\label{zset}
\eeq
where, as above, the symbol $\{G\}$ denotes the limit of the graph family $G$
as $n(G) \to \infty$ (and the $z$ here should not be confused with the 
auxiliary expansion 
variable in the generating function (\ref{gammatfbc}) or the Potts partition 
function $Z(G,q,v)$.)  General inequalities for these were given in \cite{a}. 

Our results yield 
\beq
z_{ST}(\{G\}) = \ln2+ \frac{1}{2}\ln 3 \simeq 1.242453.. 
\quad {\rm for} \quad G=S,L
\label{zst}
\eeq
\beq
z_{SF}(\{G\}) = \frac{1}{2}\ln(9+5\sqrt{3} \ ) \simeq 1.435658 
\quad {\rm for} \quad G=S,L
\label{zsf}
\eeq
\beq
z_{CSSG}(\{G\}) = \frac{1}{2}\ln [2(7+\sqrt{46} \ )] \simeq 1.658267
\quad {\rm for} \quad G=S,L
\label{tl12asymp}
\eeq
and
\beq
z_{SSG}(\{G\}) = \frac{5}{2}\ln 2 \simeq 1.732868 
\quad {\rm for} \quad G=S,L
\label{zssgl}
\eeq

In Table I we summarize the results for $z_{s}(\{G\})$ for the infinite-length
limit of the width $L_y=2$ strips of the square, triangular, and $sq_d$
strips (which are independent of the longitudinal boundary conditions).  In
this table we include a comparison of the exact values of $z_{ST}$ that we have
calculated with the upper bound (u.b.) for a $k$-regular graph \cite{mckay}
\beq
z_{ST} < z_{ST,u.b.} \ , \quad z_{ST,u.b.} = 
\ln \Biggl ( \frac{(k-1)^{k-1}}{[k(k-2)]^{(k/2)-1}} \Biggr ) 
\label{zmckay}
\eeq
The cyclic $L_y=2$ strips of the square, triangular, and $sq_d$ lattices are
$k$-regular, with $k=3,4,5$, respectively.  One thus has
\beq
z_{ST,u.b.}(sq(L_y=2)) = \ln(4/\sqrt{3})
\label{zuppersq}
\eeq
\beq
z_{ST,u.b.}(tri(L_y=2)) = 3\ln(3/2)
\label{zuppertri}
\eeq
\beq
z_{ST,u.b.}(sq_d(L_y=2)) = 8\ln 2 - \frac{3}{2}\ln 15
\label{zuppersqd}
\eeq
For this table we define 
\beq
r_{ST}(\{G\}) = \frac{z_{ST}(\{G\})}{z_{ST,u.b.}(\{G\})} \ . 
\label{rst}
\eeq
As is evident from the table, $z_{ST}(\{G\})$, $z_{SF}(\{G\})$, and 
$z_{CSSG}(\{G\})$ increase as one goes from the $L_y=2$ strip of the square
strip to that of the triangular, and $sq_d$ strip, as the degree increases from
3 to 4 to 5.  This also follows from (\ref{nssglmg}) for $z_{SSG}$.  A similar
dependence on degree (coordination number) was found for $z_{ST}$ in 
\cite{wu77,st}. 

\begin{table}
\caption{\footnotesize{$z_s(\{G\})$ for the $L_x \to \infty$ limit of the width
$L_y=2$ strips of the (i) square lattice, (ii) triangular lattice (constructed
by starting with a square strip and adding edges joining the lower left to
upper right vertex of each square, and (iii) $sq_d$ lattice, i.e., square
lattice with next-nearest-neighbor couplings. }}
\begin{center}
\begin{tabular}{|c|c|c|c|}
\hline
\hline
$z_s(\{G\})$      & $\{G\}=sq$  & $\{G\}=tri$  & $\{G\}=sq_d$ \\
\hline
$z_{ST}(\{G\})$   & $(1/2)\ln(2+\sqrt{3} \ )$
                  & $(1/2)\ln[(7+3\sqrt{5} \ ) /2]$ 
                  & $\ln(2\sqrt{3} \ )$ \\ 
                  & $=0.658479$ & $=0.962424$ & $=1.242453$ \\ 
\hline
$r_{ST}(\{G\})$   & 0.7867 & 0.7912 & 0.8377 \\ 
\hline
$z_{SF}(\{G\})$   & $(1/2)\ln[2(2+\sqrt{3} \ )]$
                  & $(1/2)\ln[2(3+2\sqrt{2} \ )]$
                  & $(1/2)\ln(9+5\sqrt{3} \ )$ \\
                  & $=1.005053$ & $=1.227947$ & $=1.435658$ \\ 
\hline
$z_{CSSG}(\{G\})$ & $(1/2)\ln[(5+\sqrt{17}\ )/2]$ 
                  & $(1/2)\ln[2(3+2\sqrt{2} \ )]$
                  & $(1/2)\ln[2(7+\sqrt{46} \ )]$ \\
                  & $=0.758832$ & $=1.227947$ & $=1.658267$ \\
\hline
$z_{SSG}(\{G\})$  & $(3/2)\ln 2$
                  & $2\ln 2$
                  & $(5/2)\ln 2$ \\
                  & $1.039721$ & $=1.386294$ & $=1.732868$ \\
\hline
\end{tabular}
\end{center}
\label{table1}
\end{table}

\vfill
\eject

\end{document}